\documentclass[fleqn,usenatbib]{mnras}

\usepackage[usenames,dvipsnames]{xcolor}

\hypersetup{
    colorlinks = true,
    citecolor = {MidnightBlue},
    linkcolor = {BrickRed},
    urlcolor = {BrickRed},
}

\usepackage{newtxtext,newtxmath}

\usepackage[T1]{fontenc}
\usepackage{ae,aecompl}
\usepackage{multirow}
\allowdisplaybreaks[1]

\usepackage{graphicx}	
\usepackage{amsmath}	
\usepackage{amssymb}	
\usepackage{bm}
\usepackage{balance}

\usepackage{geometry}
\renewcommand{\mathbf}[1]{{\bm #1}}

\newcommand{\norm}{a_{\rm N}}

\newcommand\lsim{\mathrel{\rlap{\lower4pt\hbox{\hskip1pt$\sim$}}
        \raise1pt\hbox{$<$}}}
\newcommand\gsim{\mathrel{\rlap{\lower4pt\hbox{\hskip1pt$\sim$}}
        \raise1pt\hbox{$>$}}}

\newcommand{\bn}{\mathbf{n}}

\newcommand{\bV}{\mathbf{V}} 
\newcommand{\bx}{\mathbf{x}}

\newcommand{\bk}{\mathbf{k}}

\newcommand{\HH}{{\cal H}}

\newcommand{\vv}{{\rm v}}

\def\be{\begin{equation}}
\def\ee{\end{equation}}
\def\bea{\begin{eqnarray}}
\def\eea{\end{eqnarray}}

\title[Testing GR with Doppler magnification]{Testing General Relativity with the Doppler magnification effect}

\author[S. Andrianomena et al.]{Sambatra Andrianomena$^{1,2}$\thanks{e-mail: \href{mailto:andrianomena@gmail.com}{andrianomena@gmail.com}},
Camille Bonvin$^{3\thanks{e-mail: \href{mailto:camille.bonvin@unige.ch} {camille.bonvin@unige.ch}}}$, 
David Bacon$^{4}$, 
Philip Bull$^{5,2,6}$, 
\newauthor Chris Clarkson$^{5,2,7}$, 
Roy Maartens$^{2,4}$, 
Teboho Moloi$^{7}$ 
\\\\
$^{1}$South African Radio Astronomy Observatory (SARAO), Black River Park, Cape Town 7925, South Africa \\
$^{2}$Department of Physics \& Astronomy, University of the Western Cape, Cape Town 7535, South Africa \\
$^{3}$D\'epartement de Physique Th\'eorique and Center for Astroparticle Physics, University of Geneva,
CH-1211 Geneva, Switzerland \\
$^{4}$Institute of Cosmology and Gravitation, University of Portsmouth, Portsmouth, PO1 3FX, UK\\
$^{5}$School of Physics \& Astronomy, Queen Mary University of London, London E1 4NS, UK\\
$^{6}$Radio Astronomy Laboratory, University of California Berkeley, Berkeley, CA 94720, USA \\
$^{7}$Department of Mathematics \& Applied Mathematics, University of Cape Town, Cape Town 7701, South Africa
}

\date{Accepted 2019 July 8. Received 2019 July 8; in original form 2018 October 31}

\pubyear{2019}

\begin{document}
\label{firstpage}
\pagerange{\pageref{firstpage}--\pageref{lastpage}}
\maketitle

\begin{abstract}
The apparent sizes and brightnesses of galaxies are correlated in a dipolar pattern around matter overdensities in redshift space, appearing larger on their near side and smaller on their far side. The opposite effect occurs for galaxies around an underdense region. These patterns of apparent magnification induce dipole and higher multipole terms in the cross-correlation of galaxy number density fluctuations with galaxy size/brightness (which is sensitive to the convergence field). This provides a means of directly measuring peculiar velocity statistics at low and intermediate redshift, with several advantages for performing cosmological tests of GR. In particular, it does not depend on empirically-calibrated scaling relations like the Tully-Fisher and Fundamental Plane methods. We show that the next generation of spectroscopic galaxy redshift surveys will be able to measure the Doppler magnification effect with sufficient signal-to-noise to test GR on large scales. We illustrate this with forecasts for the constraints that can be achieved on parametrised deviations from GR for forthcoming low-redshift galaxy surveys with DESI and SKA2. Although the cross-correlation statistic considered has a lower signal to noise than RSD, it will be a useful probe of GR since it is sensitive to different systematics.\\ 
\end{abstract}



\section{Introduction}

Until recently, the application of General Relativity (GR) to cosmology has represented a tremendous extrapolation of the theory to distance scales far exceeding those over which it has been subjected to precision tests. The most stringent tests of GR remain those involving experiments in the Solar System \citep[e.g. with the Cassini probe, lunar laser ranging, and Earth-orbit frame dragging and equivalence principle experiments;][]{Bertotti:2003rm, Williams:2004qba, Will:2005va, Everitt:2011hp, Touboul:2017grn}, and observations of binary pulsar systems \citep{Taylor:1982zz, Taylor:1991yt, EspositoFarese:1996si, Weisberg:2004hi, kramer2006tests, Wex:2014nva}, all covering distances substantially less than a parsec. 

A host of new precision tests are starting to become feasible that can greatly extend the range over which GR has been validated, however \citep{Damour:1991rd, Baker:2014zba, Berti:2015itd, Sakstein:2017pqi}. A notable example is the recent detection of gravitational waves from binary black hole and neutron star coalescences by the LIGO and VIRGO detectors \citep{Abbott:2016blz, Monitor:2017mdv}. All of the events that have been observed so far appear to be consistent with GR \citep{TheLIGOScientific:2016src}, thus extending a subset of precision tests out to a comoving distance of $\sim 800$ Mpc ($z \simeq 0.2$) for the most distant event seen so far \citep{Abbott:2017vtc}. This is sufficient to extend the envelope of precision tests out into the Hubble flow, beyond the gravitational environment dominated by our local cluster.

On larger scales and at higher redshifts, an extremely wide variety of tests have been proposed, involving such diverse objects and observables as galaxy clusters and their mass function; supermassive black holes embedded in galaxies; weak lensing distortions of galaxies, clusters, and the CMB; and the redshift-space clustering of galaxies, including relativistic effects~\citep{Bonvin:2018ckp}; see \cite{Berti:2015itd} for a review. These cover a broad range of distance scales and gravitational environments, and can be quite precise in some instances \citep[e.g. for Chameleon models;][]{Burrage:2016bwy}. A number of technical issues stand in the way of attaining the comprehensive, high-precision constraints that have been achieved in the Solar System however. First, most of these tests depend on accurately modelling complex astrophysical phenomena, which introduces significant systematic uncertainties. Second, astronomical measurements are inherently noisier, and require considerably more data to reach similar levels of precision to Solar System tests. Finally, possible deviations from GR are more diverse and harder to parametrise in the cosmological regime \citep[c.f. the Parametrized Post-Newtonian framework on Solar System scales;][]{2011PNAS..108.5938W}, leaving many tests effectively model-dependent.

To overcome these difficulties, cosmological tests of GR are needed that are sufficiently sensitive and general while being less susceptible to astrophysical systematics. A particularly promising class of observables involve direct measurements of the peculiar velocity field \citep[e.g.][]{2009JCAP...10..017K, 2012ApJ...751L..30H, 2014PhRvL.112v1102H, 2015ApJ...808...47M, 2015MNRAS.449.2837G, 2016MNRAS.458.2725J, 2016A&A...595A..40I}. Velocities are a sensitive probe of gravitational physics, as they respond to changes in the effective strength of gravity over long periods of time. Since most non-GR theories are expected to modify the growth rate of structure, they should therefore leave an imprint in the cosmic peculiar velocity distribution. 
The equivalence principle implies that freely-falling galaxies should all respond to gravitational potentials in the same way, regardless of their mass or type, so galaxy peculiar velocities should also be unbiased with respect to the underlying dark matter distribution \citep[at least to linear order on large scales; see][]{2015PhRvD..91l3512Z, 2016arXiv161109787D}. This means that velocities do not depend on tracer-dependent bias terms, which are an important source of uncertainty for other galaxy clustering observables -- particularly as they can be degenerate with signatures of modified gravity \citep[e.g.][]{2014MNRAS.440...75B, 2014JCAP...04..029B}. Combinations of observables that have similar bias-independent properties can be constructed in principle, such as the $E_{\rm G}$ statistic that combines galaxy density and lensing measurements \citep{2007PhRvL..99n1302Z, 2010Natur.464..256R}, but tracer-dependent quantities tend to re-enter the resulting quantity in practice \citep{2015JCAP...12..051L,Dizgah:2016bgm}.

While this makes direct velocity-based observables cleaner in principle, most practical methods of measurement reintroduce dependences on hard-to-model astrophysical phenomena. The Tully-Fisher method \citep{1977A&A....54..661T} commonly used at low redshifts relies on an empirically-calibrated scaling relation between the luminosity and circular velocity of a galaxy, for example. Similarly, the fundamental plane and Faber-Jackson relations, which are used to measure velocities from elliptical galaxies, are constructed from empirical relations between the luminosity and stellar velocity dispersions~\citep{Faber:1976sn, 1987ApJ...313...59D}. Furthermore, the kinetic Sunyaev-Zel'dovich effect can be used to measure the velocities of galaxy clusters at higher redshifts, but only in a way that is degenerate with the integrated optical depth of the cluster, which must be modelled based on other measurements \citep{2008JCAP...08..030B, 2015ApJ...808...47M, Alonso:2016jpy, Battaglia:2016xbi}. Nevertheless, these methods have all been successfully used to make cosmological measurements in the past, including some that even appear to show deviations from $\Lambda$CDM+GR \citep[e.g.][]{Kashlinsky:2008ut, 2009MNRAS.392..743W, 2011MNRAS.414..621M}. Concerns about astrophysical systematics have contributed to scepticism of these anomalous results, however, which remain contentious.

In this paper, we describe how tests of GR can be performed using a different peculiar velocity observable called Doppler magnification. Doppler magnification concerns the effect of peculiar velocities on the apparent sizes of objects in redshift-space. As shown in~\cite{Bonvin:2008ni}, a galaxy with a component of its peculiar velocity directed away from us is physically closer to us than a galaxy at the same redshift with no peculiar velocity. As a consequence, the moving galaxy appears larger than the one with no velocity, which is akin to a magnification of the moving galaxy. Conversely a galaxy with a peculiar velocity directed towards us is physically further away, which leads to its apparent demagnification. The Doppler magnification contributes therefore to the convergence, in addition to the standard gravitational lensing contribution: for galaxies at low redshift, $z \leq 0.5$, Doppler magnification dominates, whereas for high redshift galaxies, gravitational lensing dominates~\citep{Bonvin:2008ni}. As shown in~\cite{Bacon:2014uja}, detecting galaxy peculiar velocities from the convergence auto-correlation will be very challenging. However by cross-correlating the convergence with the galaxy density field, one can significantly enhance the signal and reach a detectable level. In particular \cite{Bacon:2014uja} forecasted the constraints on cosmological parameters that can be obtained with the density-convergence correlation, and showed that Doppler magnification is a competitive large-scale structure observable. \cite{Bonvin:2016dze} subsequently showed that the optimal way to measure the density-convergence correlation is to fit for a dipole and octupole. This can be intuitively understood by noting that galaxies around an overdensity tends to move toward it, which systematically magnifies galaxies in front of the overdensity and demagnifies galaxies behind. \cite{Bonvin:2016dze} furthermore showed that the dipole and octupole are dominated by Doppler magnification up to high redshift, of order 1. In this paper, we build on these previous studies, by showing that the Doppler magnification dipole can be used to test GR.  

This is similar in spirit to methods like the Tully-Fisher effect, except we use a `statistical ruler' (galaxy sizes) rather than standard candles (see \cite{Kaiser:2014jca} for a discussion). The need to model the properties of the target galaxy population is also much reduced compared to these methods. There is no need to separately calibrate an empirical scaling relation for example, as the mean galaxy size as a function of redshift can be measured from the survey itself. Also, the galaxy bias does not enter into the velocity-velocity term, which can be measured directly from the octupole of the correlation function. Measurement systematics do of course remain (the size estimates can depend on sensor characteristics, for example), but can at least be mitigated through experimental design or high-fidelity instrumental simulations and calibration strategies. This goes some way to achieving what the $E_{\rm G}$ statistic originally sets out to do -- removing the bias dependence whilst also being sensitive to modified gravitational physics. Our method slightly differs from Tully-Fisher measurements however, due to the fact that our estimator is sensitive not only to the impact of peculiar velocities on the galaxy sizes, but also to their impact on the galaxy density field. It is therefore a hybrid method, combining direct velocity observables with redshift-space distortions (RSDs). An immediate consequence of this is that Doppler magnification depends on both the velocity field and its gradient, whereas RSDs are only generated by the gradient of the velocity. As discussed in Section~\ref{sec:mgdoppler}, this different behaviour implies a different sensitivity of these two probes to modified gravity models, especially those with a growth of structure that depends on scale. Doppler magnification is therefore complementary to RSDs by construction as a probe of GR. Its only drawback is that, due to uncertainties in the size measurements, it will be measured with a considerably lower signal-to-noise than RSDs.

The paper is organised as follows. In Section~\ref{sec:mgdoppler}, we review the Doppler magnification effect and how it can be extracted from the cross-correlation of the galaxy density and convergence fields, and show how the relevant quantities are affected by deviations from GR. We present forecasts for the detectability of deviations from GR with DESI and SKA HI spectroscopic galaxy surveys in Section~\ref{sec:forecasts}, and then conclude in Section~\ref{sec:conclusions}.

We use a flat $\Lambda$CDM+GR cosmology with cosmological parameters $h=0.68$, $\Omega_{\rm cdm} = 0.2548$, $\Omega_b=0.048$, $n_s=0.96$, and $\sigma_8 = 0.83$ as the fiducial model. 

\section{Doppler magnification in modified gravity} \label{sec:mgdoppler}

In this section we briefly review the Doppler magnification effect, and a method to extract it from the dipole of the number count-convergence correlation function. We then discuss how modifications to GR affect the Doppler dipole, and study some illustrative examples of modified gravity theories, and their effects on the number count-convergence correlation.

\subsection{Number count-convergence correlation} \label{sec:correlation}

\begin{figure}
\centering
\includegraphics[width=0.5\columnwidth]{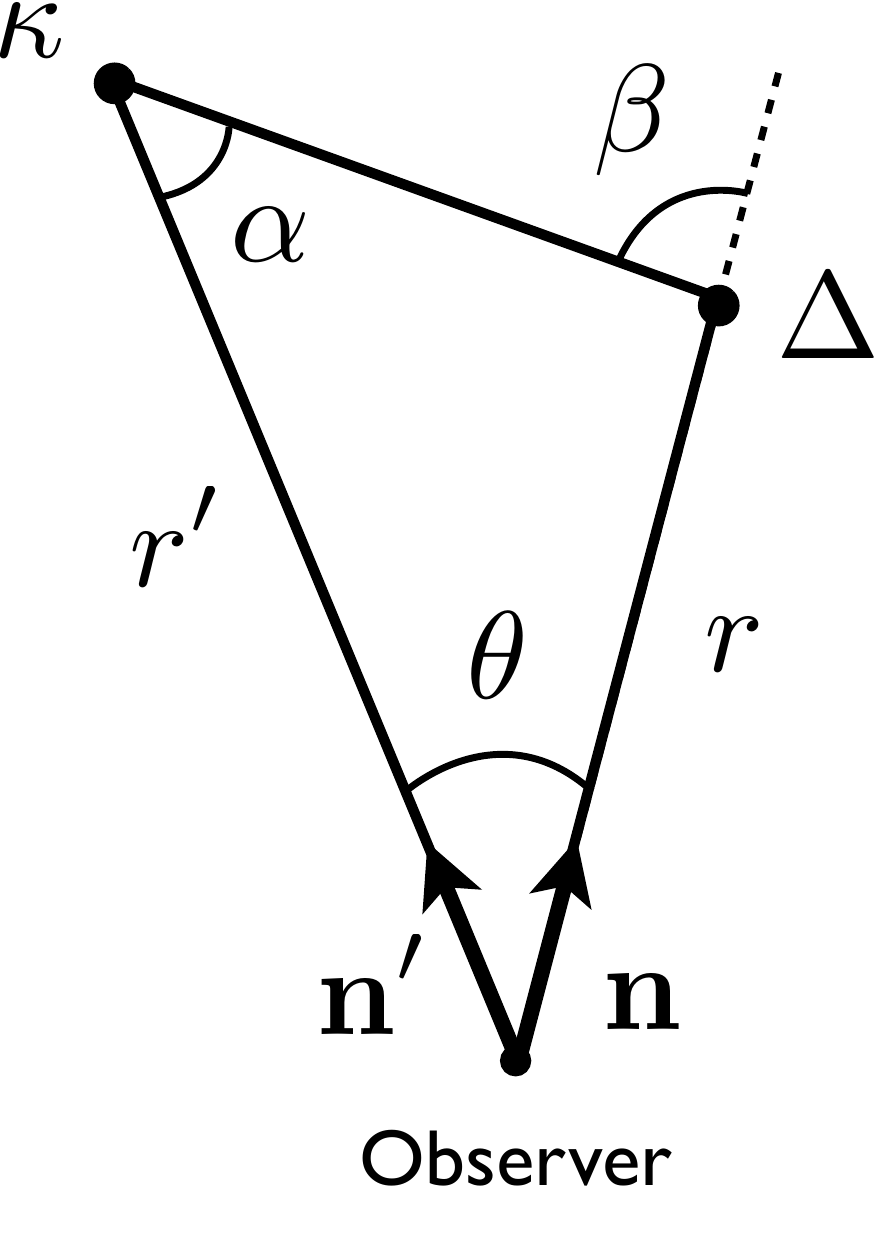}
\caption{The geometrical set-up and definitions of various angles used in the calculations. }
\label{coorsystem}
\end{figure}

The Doppler magnification effect is observed in the cross-correlation of the galaxy number count fluctuation, $\Delta$, and a suitable proxy for the convergence field, $\kappa$, which can be constructed by combining size and magnitude measurements~\citep{Schmidt:2011qj,Casaponsa:2012tq,Heavens:2013gol,Alsing:2014fya}. We begin by calculating the cross-correlation as a function of redshift and angle,
\be\label{xi}
\xi^{\Delta \kappa} = \langle \Delta(z,\bn)\kappa(z',\bn') \rangle\, ,
\ee
where $\bn$ denotes the direction of observation.
This quantity can be expanded in a hierarchy of multipoles around an observed overdensity, with various correlations between the density, velocity, and lensing terms contributing differently to each multipole as a function of redshift.

At linear order, the galaxy number count fluctuation is given by
\be
\label{numbercount}
\Delta(z, \bn) \simeq b\,\delta-\frac{1}{\HH}\partial_r(\bV\cdot\bn)\,,
\ee
where $b$ is the local bias, $\HH \equiv a H$ is the conformal Hubble rate, $r$ is the comoving distance of the galaxy and $\bV\cdot\bn$ is the line-of-sight peculiar velocity of the galaxy. The second term is the redshift-space distortion. We have neglected here the lensing and relativistic corrections to $\Delta$~\citep{Yoo:2009au,Bonvin:2011bg,Challinor:2011bk} since their contribution to the cross-correlation is subdominant at low redshift and on sub-horizon scales.\footnote{The contribution from the relativistic corrections to the dipole and octupole of $\langle\Delta\kappa\rangle$ is suppressed by $(d/r)^2$ ($d$ being the pixels' separation) with respect to the contribution from the standard Newtonian terms in Eq.~\eqref{numbercount}. The lensing contribution is strongly subdominant at small redshift, similar to what was found in~\cite{Bonvin:2013ogt}.}

The convergence contains two dominant contributions: the standard weak lensing convergence, $\kappa_{\rm g}$, and a Doppler contribution~\citep{Bonvin:2008ni,Bacon:2014uja}
\be\label{kappav}
\kappa_\vv \equiv \left(\frac{1}{r\HH} -1\right)\bV\cdot\bn\, .
\ee
Suppose that we look at a galaxy with a peculiar velocity directed towards the observer, such that $\bV\cdot\bn<0$. This galaxy is physically further away than a galaxy with the same redshift and no peculiar velocity. The first term in Eq.~\eqref{kappav} is then negative, leading to a demagnification of the galaxy. This term simply reflects the fact that a galaxy which is further away appears smaller. The second term however has the opposite sign. It is due to the fact that a galaxy which is more distant is situated at a smaller value of the scale factor. It experiences therefore a larger apparent stretch, due to the expansion of the Universe between emission and observation and appears larger. At small redshift the first term dominates, whereas at large redshift the second one dominates. These two terms compensate around $z\simeq 1.6$, leaving the size of the galaxy unchanged.
Comparing the Doppler term, with the gravitational lensing term, we see that the first one dominates at low redshift $z\lsim 0.5$, whereas the lensing term dominates at high redshift~\cite{Bonvin:2008ni}. 

Note that the convergence contains also a contribution from the peculiar velocity at the observer. This term generates a local dipole, which can be subtracted from the data. In any case, its contribution to the number count-convergence correlation is expected to be completely negligible because it affects the size of all galaxies around a given overdensity in almost the same way, and it does therefore not contribute to a dipolar modulation around an overdensity (except when the separation becomes of the order of the comoving distance $r$).

As shown in~\cite{Bonvin:2016dze}, the correlation of $\kappa_\vv$ with $\Delta$ has a distinctive dipolar structure. For a realistic survey, this can be projected out and measured using an estimator of the form~\citep{Bonvin:2016dze}
\be \label{eq:ddipole}
\xi_{\rm dip}(d)=\norm \sum_{ij} \Delta_i \kappa_j \cos\beta_{ij}\delta_K(d_{ij}-d),
\ee 
where $\norm$ is a normalisation factor, and the sum is over pairs of pixels in the survey separated by a physical comoving distance $d_{ij}$. The angle $\beta_{ij}$ is the angle formed at $\Delta_i$ between  the line-of-sight angle $\bn$ and the direction vector to $\kappa_j$ (see Fig.~\ref{coorsystem}). In what follows, we will refer to the quantity in Eq.~\eqref{eq:ddipole} as the `Doppler magnification dipole'. This estimator provides an efficient way of isolating the $\kappa_\vv$ contribution. Projecting $\Delta\kappa$ onto a dipole does indeed strongly suppress the gravitational lensing contribution $\kappa_{\rm g}$ up to redshift $\sim 1$, see~\cite{Bonvin:2016dze}. Note that the cross-correlation between $\Delta$ and $\kappa$ also contains an octupole modulation, which can be isolated by weighting the two-point function by the Legendre polynomial $P_3(\cos\beta_{ij})$. As shown in~\cite{Bonvin:2016dze}, this octupole is however significantly smaller than the dipole, and we therefore concentrate only on the latter in the following.

As implied by Eqs.~(\ref{kappav}) and (\ref{eq:ddipole}), the Doppler dipole can be used as a probe of the line-of-sight peculiar velocity field. This is of particular interest in studies of modified gravity theories, which generically alter the growth rate of structure, $f$. Since at linear order we have $\bV \propto f$, this suggests that Doppler magnification can be used as an independent probe of gravitational physics on cosmological scales. Note that the correlation $\langle\Delta\kappa\rangle$ depends on $f$ in a different way than the standard redshift-space distortion terms in the $\langle\Delta\Delta\rangle$ correlation, since $\kappa \propto \bV\cdot\bn$, whereas $\Delta$ depends on the velocity gradient $\partial_r (\bV\cdot\bn)$. As such the sensitivity of these two probes to modified gravity may be different, especially in the case where the growth rate is scale-dependent. Doppler  magnification  is  therefore expected to be highly  complementary  to  RSDs  for  probing  GR.  One drawback of this probe is however that, due to difficulties in measuring the size of galaxies, its signal-to-noise will be significantly lower than that of RSDs.

In the following sections, we show how deviations from General Relativity enter into the calculation of the Doppler magnification dipole, and derive expressions for the dipole (and higher multipoles) of the number count-convergence correlation function in the presence of such effects.

\subsection{Velocity potential and growth factor}

While alternative theories of gravity can be extremely complex in general, the vast majority can be described by a handful of new functional degrees of freedom to linear order in perturbations on an assumed FLRW background with an effective dark energy equation of state $w(z)$ \citep{Amendola:2012ky, Baker:2012zs, Gleyzes:2013ooa, Lagos:2017hdr}. 
A further simplification can be made by applying the {\it quasi-static approximation}, which neglects time derivatives of any new gravitational degrees of freedom, and by restricting our attention to scales much smaller than the horizon ($k \gg \mathcal{H}$).
Under these assumptions, we follow the common practice (e.g.\ \citep{Pogosian:2010tj}) and define the Poisson-like equation relating  the  density  and  time-time  gravitational  potential in Fourier space as
\footnote{We assume the following metric convention throughout this paper: $ds^{2} = a^{2}\left[-(1+2\Psi)d\tau^{2}+(1-2\Phi) \gamma_{ij}dx^{i}dx^{j}\right]$, and we use the Fourier convention $f(\mathbf{x}, \tau)=(2\pi)^{-3} \int d^3\bk e^{-i\bk\cdot\bx}f(\bk,\tau)$.}
\be \label{poiss}
-k^{2}\Psi = 4\pi G a^{2}\bar{\rho}\mu(a,k)\,\delta,
\ee
where $\mu(a,k)$ is an arbitrary function of scale factor and wavenumber that encodes the modified gravitational physics, and has the value $\mu=1$ in GR.
A second modification also arises, in the form of a non-trivial `gravitational slip' relation relating the two potentials in the metric, $\Phi = \eta(a,k)\Psi$. 
The slip parameter $\eta$ is again an arbitrary function of time and scale, and $\eta=1$ in GR. 
Finally, theories of modified gravity can break Einstein's equivalence principle, generating modifications to Euler's equation \cite[see][]{Gleyzes:2015pma}. We do not consider this possibility here and we assume that Euler's and the continuity equations are the same as in GR.

The growth equation for $\delta$ is determined by solving the modified Bardeen equation on sub-Hubble scales \citep{Pogosian:2010tj,Amendola:2012ys},
\be \label{bardnmg}
{\partial^2\delta \over \partial (\ln a)^2} + \left(2+\frac{\partial \ln H}{\partial \ln a}\right){\partial \delta\over \partial \ln a} = \frac{3}{2}\left(\frac{H_0}{H}\right)^2\frac{\Omega_{m}}{a^3}\mu\delta\,,
\ee
where $\Omega_m$ denotes the matter density parameter today. From Eq.~\eqref{bardnmg} we see that $\eta$ does not enter into the growth equations for $\delta$ on sub-Hubble scales. Moreover, since $V$ is directly related to $\delta$ by the continuity equation (Eq.~\eqref{eq:contuity} below), the growth of $V$ is also independent of $\eta$ (see also the evolution equation for $V$, Eq. (38) in~\cite{Hall:2012wd}). Since $\delta$ and $V$ are the only quantities contributing to the Doppler magnification dipole in the regime we are interested in (where the lensing contribution is negligible), our observable will be insensitive to $\eta$. 
Let us however mention that this is specific to the choice of parameterisation chosen here. In GR, the Poisson equation relates in fact the density perturbation $\delta$ to the spatial component of the metric, i.e.\ the potential $\Phi$. In Eq.~\eqref{poiss} we have modified the Poisson equation, such that $\mu$ relates $\delta$ to the time component of the metric, namely $\Psi$. Hence $\mu$ encodes both a deviation in the growth of structure, and a difference between the two gravitational potentials.

We write the solution to Eq.~\eqref{bardnmg} in Fourier space in terms of the usual $\Lambda$CDM transfer function $T(k)$ and the primordial scalar potential $\Psi_{\rm p}({\bm k})$, such that 
\be \label{eq:poisson2}
\delta({\bm k},z) = -\frac{2}{3}\left(\frac{k}{\mathcal{H}_{0}}\right)^{2}\frac{T(k)D(z,k)}{\Omega_{m}\mu(z,k)}\Psi_{\rm p}({\bm k}).
\ee
The growth factor $D$ is found by solving Eq.~\eqref{bardnmg} assuming that at early times, deep in the matter era, density perturbations grow linearly: {$D(k, a_{\rm ini})=a_{\rm ini}$, and $[dD(k, a)/da]_{\rm ini}=1$}.
Inserting~\eqref{eq:poisson2} into the continuity equation on sub-Hubble scales 
\be
\label{eq:contuity}
V({\bm k},z)=-\frac{1}{k}
\dot{\delta}({\bm k},z)\,,
\ee
where 
a dot denotes a conformal time derivative, the velocity potential becomes
\be \label{velmg2}
V({\bm k},z) = \mathcal{G}(k,z) T(k)\Psi_{\rm p}({\bm k})\, .
\ee
The `velocity growth factor' $\mathcal{G}$ is defined by 
\be\label{growth_G}
\mathcal{G}(a, k) \equiv \frac{2ak}{3\mu \mathcal{H}_{0}^{2}\Omega_{m}}\left[\left({D\over a} \right)^{\displaystyle\cdot} + \frac{D}{a}\left(\mathcal{H}-\frac{\dot\mu}{\mu}\right)\right]\,.
\ee

\subsection{Multipoles of the correlation function}

Combining $\Delta$ with the Doppler convergence $\kappa_\vv$ in~Eq.~\eqref{kappav} we obtain
\bea \label{cross}
&& \xi^{\Delta \kappa_{\vv}}(z,z',\theta) = \left(\frac{1}{\mathcal{H}(z')r(z')}-1\right) \\
&& ~~~~~~ \int\frac{{\rm d}^{3}{\bm k}}{(2\pi)^{3}} e^{i{\bm k}\cdot ({\bm x'}-{\bm x})} \mathcal{G}(z^\prime, k)T^2(k)\mathcal{P}(k) i(\hat{\bm k}\cdot \hat{\textbf n}') \nonumber\\
&& ~~~~~~ \times\Biggl[\frac{2b}{3\Omega_{m}}\left(\frac{k}{\mathcal{H}_{0}}\right)^{2}\frac{D(z,k)}{\mu(z,k)} 
 +(\hat{\bm k}\cdot \hat{\textbf n})^{2}\frac{k}{\mathcal{H}(z)}\mathcal{G}(z, k)\Biggr] \,,\nonumber
\eea
where $\mathcal{P}(k)$ denotes the primordial power spectrum $\langle\Psi_{\rm p}(\bk)\Psi_{\rm p}(\bk')\rangle=(2\pi)^3 \mathcal{P}(k)\delta^{D}(\bk+\bk')$. 
Note that the primes on redshift and direction in Eq.~\eqref{cross} refer to the pixel where $\kappa$ is estimated. The cross-correlation~\eqref{cross} is a function of  $\theta$ which is the angle between $\bn$ and $\bn'$. We can re-express this cross-correlation in terms of $(z, d, \beta)$, where $d$ is the comoving distance between the galaxies and $\beta$ is the orientation of the pair with respect to the line-of-sight (see Fig.~\ref{coorsystem}). Following~\citet{Szalay:1997cc, Szapudi:2004gh, Papai:2008bd, Montanari:2012me, Bonvin:2016dze}, we expand the exponential and factors of $\hat{\bm k}\cdot\bm n$ in terms of spherical harmonics, which allows us to integrate over the direction of $\bk$. The cross-correlation then takes the simple form
\begin{align}\label{dipolefullsky}
&\xi^{\Delta \kappa_{\vv}}(z,z',\theta) = \frac{1}{2\pi^{2}}\left(1-\frac{1}{\mathcal{H}(z')r(z')}\right)\times\\
&\left\{\biggl[\frac{2\nu_{1} - 3\nu_3}{10}\left ( \frac{1}{3} + \cos2\beta \right ) + \lambda_{1} \biggr] \cos{\alpha}
+ \frac{\nu_{1} + \nu_{3}}{5}{\sin \alpha\sin2\beta}\right\}\, .\nonumber
\end{align}
The angles $\alpha$ and $\beta$ are defined in Fig.~\ref{coorsystem}, and the functions $\lambda_\ell$ and $\nu_\ell$, with $\ell = 1,3$, are given by
\begin{align}
\lambda_{\ell}(d,r, \beta) &= \int {\rm d}k k^{2}j_{\ell}(kd)\, \mathcal{G}(z^\prime, k) T^{2}(k)\mathcal{P}(k) \label{lambda}\\
&\times \Biggl[\frac{2b}{3\Omega_{m}}\left(\frac{k}{\mathcal{H}_{0}}\right)^{2}\frac{D(z, k)}{\mu(z,k)}\nonumber 
 +\frac{1}{3}\frac{k}{\mathcal{H}(z)}\mathcal{G}(z, k)\Biggr]\\
\nu_{\ell}(d,r, \beta) &= \int {\rm d}k k^{2}j_{\ell}(kd)T^{2}(k)\mathcal{P}(k)\frac{k}{\mathcal{H}(z)}\mathcal{G}(z, k)\mathcal{G}(z^\prime, k)\,.\label{nu}
\end{align}
Note that the functions $\lambda_\ell$ and $\nu_\ell$ depend not only on the pixels' separation $d$, but also on $r$ and $\beta$ through the evolution of $\mathcal{G}(z', k)$ with redshift.

To extract the dipole signal from Eq.~\eqref{dipolefullsky}, we write $z'$, $r(z')$ and $\alpha$ explicitly as a function of ($d$, $\beta$, $r(z)$) and weight the cross-correlation $\xi$ by the Legendre polynomial $P_{1}(\cos\beta)$ (i.e multiplying Eq.~\eqref{dipolefullsky} by $\cos\beta$ and integrate it over $\beta$). 

Eq.~\eqref{dipolefullsky} can be simplified using the the flat-sky approximation and neglecting evolution between $z$ and $z'$. In the flat-sky approximation the line-of-sight to $\kappa$ and $\Delta$ are approximated as parallel, such that $\alpha=\beta$, see Fig.~\ref{coorsystem}. We then have ${\cos\alpha\cos2\beta} ={-\cos\beta+2\cos^3\beta}$ and ${\sin \alpha\sin2\beta} = { 2\cos\beta-2\cos^{3}\beta}$. As discussed in~\cite{Bonvin:2016dze}, corrections to the flat-sky approximation are suppressed by the factor $\left(d/r\right)^2$ in the dipole (since $d/r$ corrections contribute only to even multipoles). Similarly evolution corrections, coming from the evolution of the functions $\HH, r$ and $\mathcal{G}$ between $z$ and $z'$ can be shown to scale as $\left(d/r\right)^2$, multiplied by the second redshift derivative of the functions. With this, we find
\be\label{dipoleflatsky}
\xi^{\Delta \kappa_{\vv}} \simeq \frac{1}{2\pi^{2}}\left(1 - \frac{1}{\mathcal{H}r}\right)\Biggl[\left(\lambda_{1}+\frac{4}{15}\nu_{1}\right)P_{1}({\cos \beta})
-\frac{2}{5}\nu_{3}P_{3}({\cos \beta})\Biggr],
\ee
where $P_\ell$ denote the orthogonal Legendre polynomials, $ P_{1}(x)=x $, $ {2}P_{3}(x)=5x^{3}-3x$.
Since we have neglected the evolution of $\mathcal{G}$ between $z$ and $z'$, the functions $\lambda_\ell$ and $\nu_\ell$ are now independent of the orientation $\beta$, so that in the flat-sky approximation, the dipole signal is just given by the $P_{1}(\cos\beta)$ term in Eq.~\eqref{dipoleflatsky}. As shown in~\cite{Bonvin:2016dze}, the octupole, $P_{3}({\cos \beta})$, is also detectable in future surveys, but we do not consider it further here since its signal-to-noise is significantly smaller than that of the dipole. In the following we will compare the flat-sky approximation with the full-sky expression and show that the difference between the two is significantly smaller than the variance of the dipole at all relevant scales. However in the forecasts, we use the full-sky expression for completeness.

The dependence of the dipole on the theory of gravity is encoded in the functions $\lambda_\ell$ and $\nu_\ell$, which depend on $D$, $\mathcal{G}$ and $\mu$. Deviations from GR have two impacts on the dipole. First, they change the evolution of the dipole with redshift, via the redshift dependence of $D$, $\mathcal{G}$ and $\mu$. And second, they change the shape of the dipole as a function of $d$. The $k$-dependences of $D$, $\mathcal{G}$ and $\mu$ do indeed modify the integrals over $k$ in Eqs.~\eqref{lambda} and~\eqref{nu}, leading to a different scaling with $d$.

\begin{figure*}
	\includegraphics[width=\columnwidth]{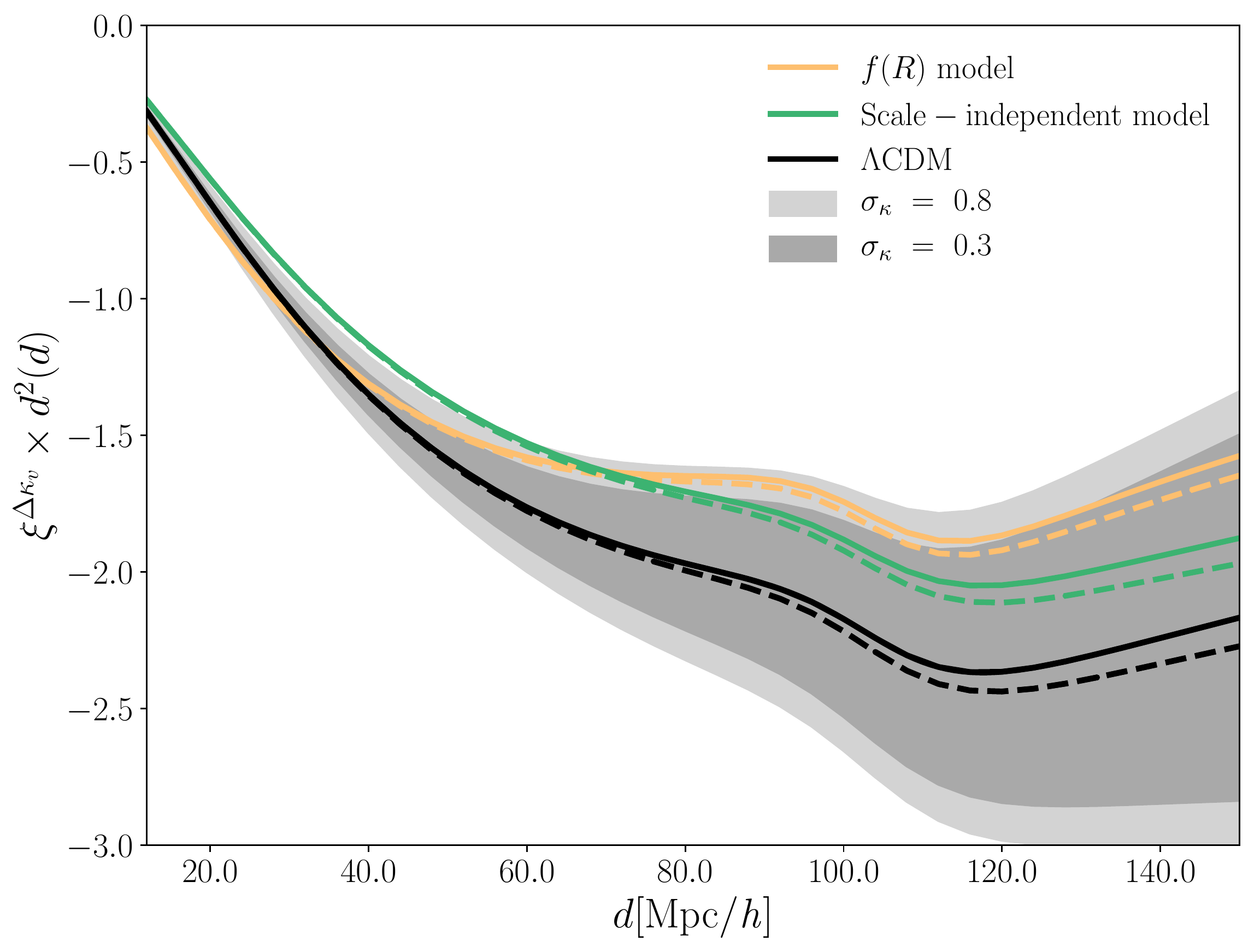}
	\includegraphics[width=\columnwidth]{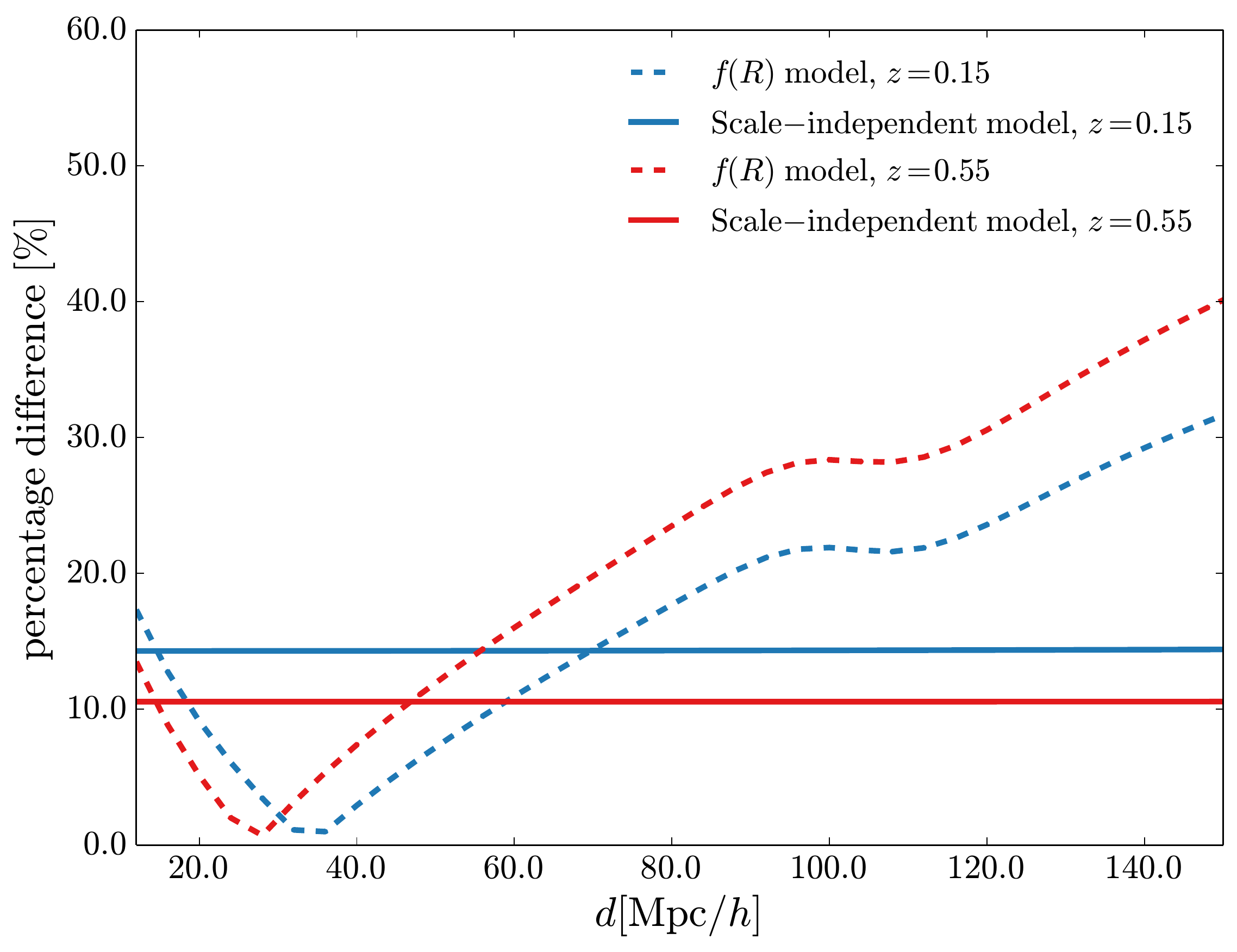}
    \caption{The survey considered here is SKA2. {\it Left:} Full-sky dipole magnification (multiplied by $d^{2}$) for $\Lambda$CDM (solid black), $f(R)$ model (solid orange) and scale-independent model (solid green) against separation $d$ at $z = 0.15$. Dashed lines are the flat-sky counterparts. 
Dark grey represents the errors when $\sigma_{\kappa} = 0.3$, light grey when $\sigma_{\kappa} = 0.8$. {\it Right:} Percentage difference between $f(R)$  and $\Lambda$CDM is shown for  $z=0.15$ (dashed blue) and $z=0.55$ (dashed red). Percentage difference between the scale-independent model and $\Lambda$CDM is shown by solid lines. In both panels we have chosen $B_0=0.1$ and $E_{11}=0.06$.}
    \label{dipole_plot}
\end{figure*}

\begin{figure}
\includegraphics[width=\columnwidth]{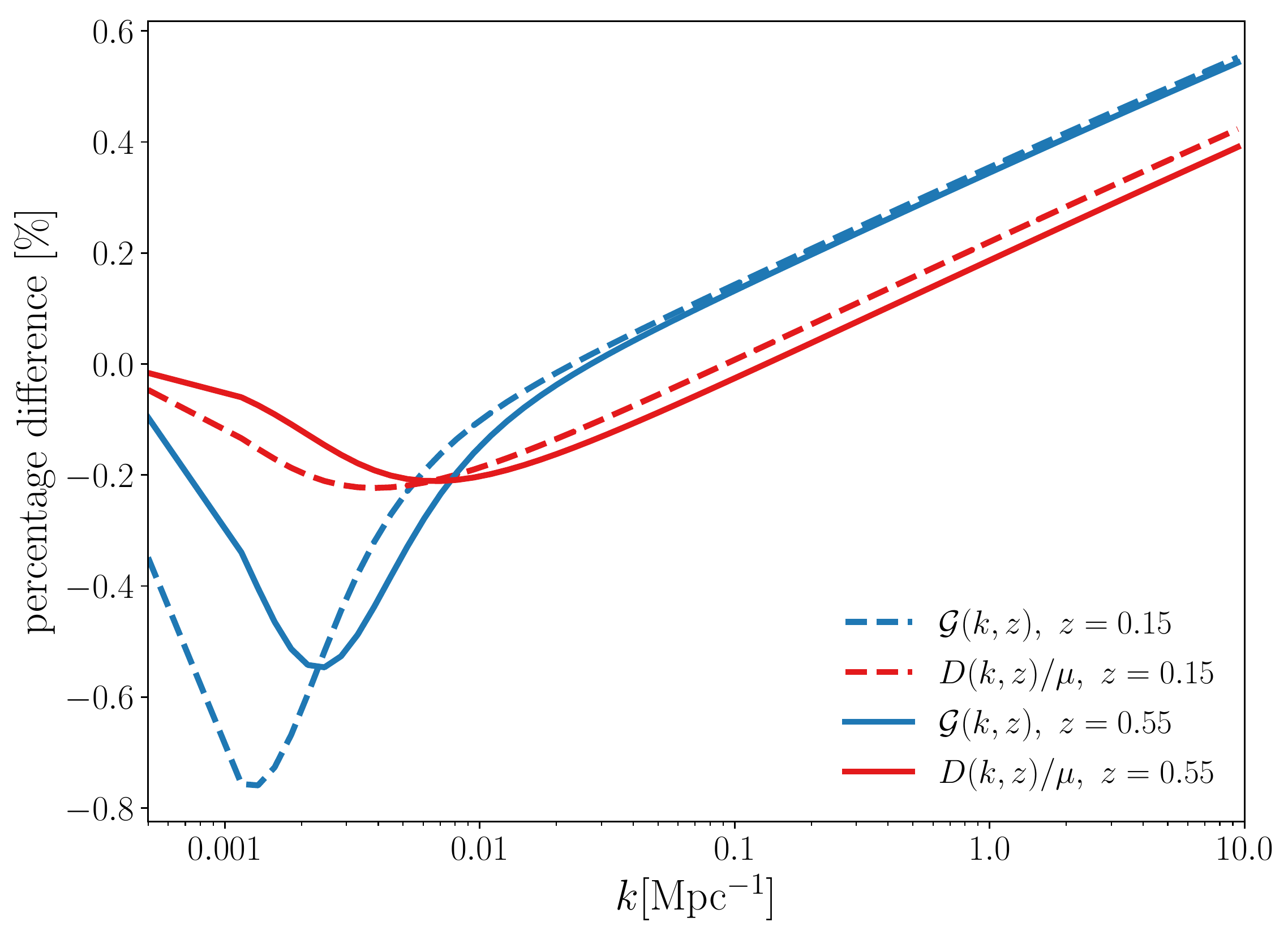}
\caption{Percentage difference in $\mathcal{G}(k,z)$ and in $D(k,z)/\mu$ between $f(R)$ and $\Lambda$CDM at $z = 0.15$ and $z = 0.55$.}
\label{fig:GDpercent}
\end{figure}

\begin{figure*}
\includegraphics[width=\columnwidth]{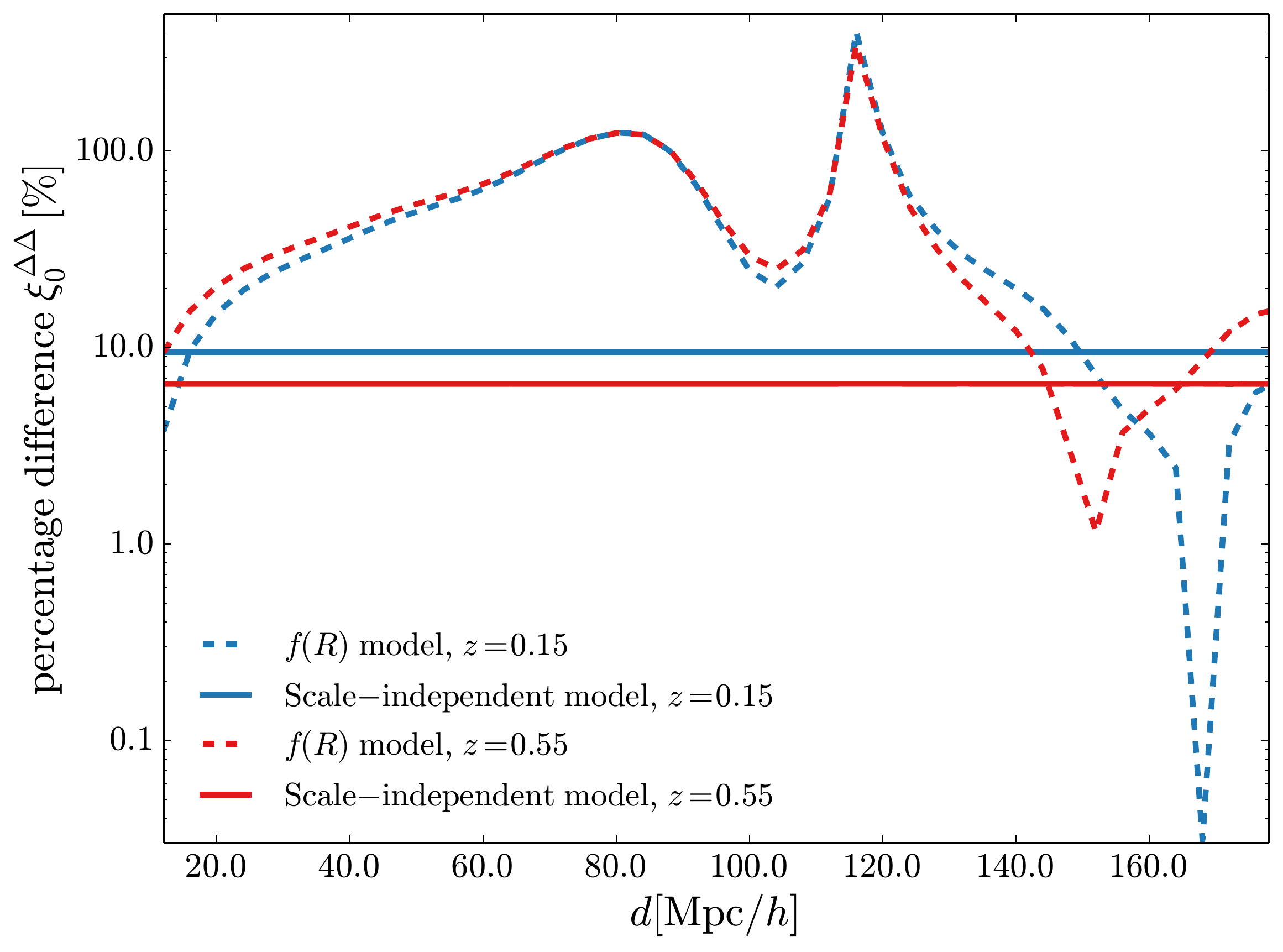}
\includegraphics[width=\columnwidth]{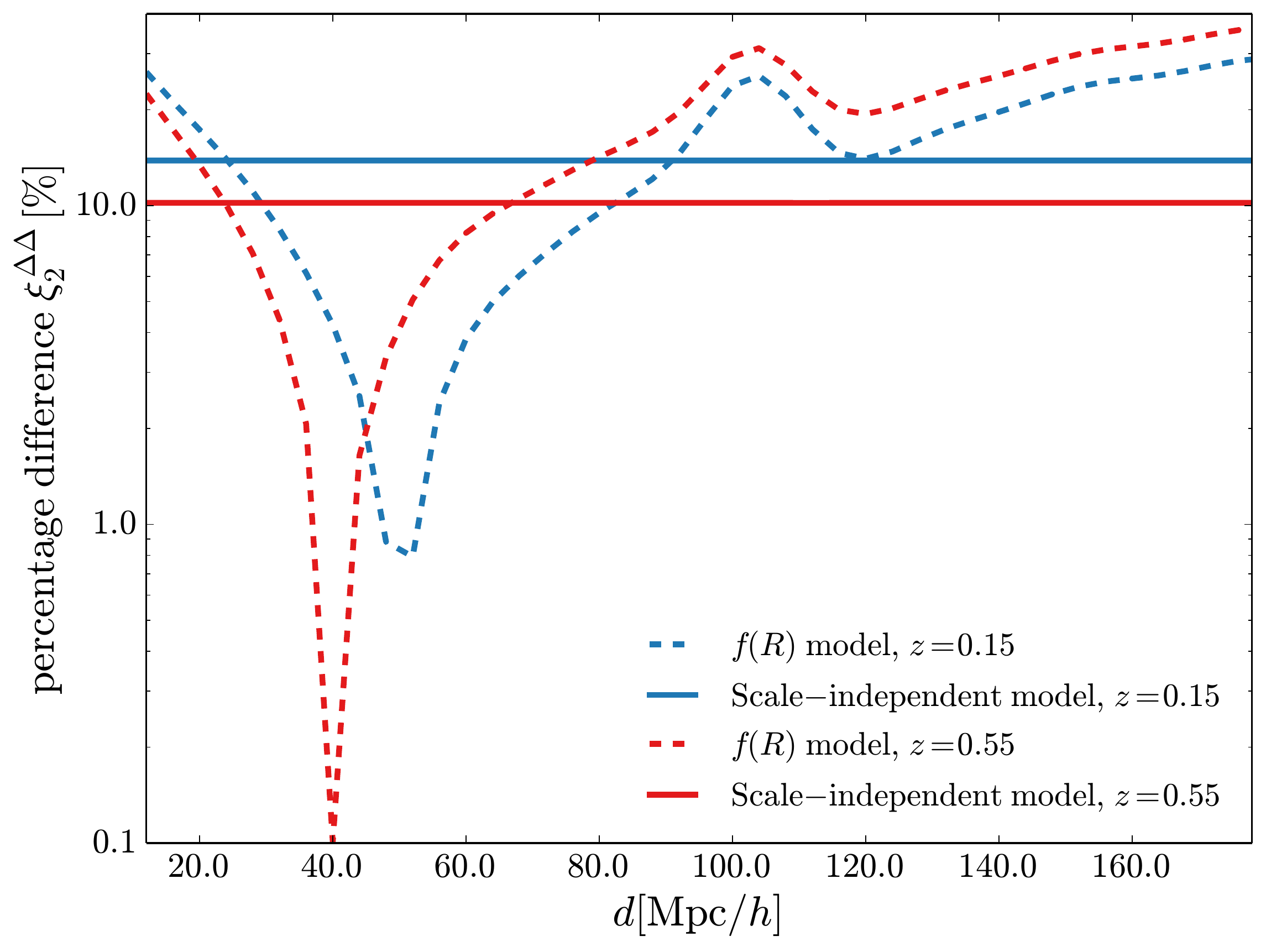}
\caption{Percentage difference in the monopole (left) and quadrupole (right) of RSD between $f(R)$ and $\Lambda$CDM is shown for  $z=0.15$ (dashed blue) and $z=0.55$ (dashed red). Percentage difference between the scale-independent model and $\Lambda$CDM is shown by solid lines. In both panels we have chosen $B_0=0.1$ and $E_{11}=0.06$. It is worth noting that the spike (around 120 Mpc/$h$) on the deviation related to the monopole is due to the fact that the two monopoles ($\Lambda{\rm CDM}$ and $f(R)$) change sign around that scale.}\label{rsd_plot}
\end{figure*}

\subsection{Modified gravity models} \label{sec:models}

We now consider two parameterisations of $\mu$ that are representative of some alternative theories of gravity. First we consider a specialisation of the model presented in~\cite{Ade:2015rim}
\be\label{debasedmu}
\mu(a,k) = 1+E_{11}\Omega_{\rm DE}(a)\, .
\ee
In this model, the growth is scale-independent, making this also an effective dark energy parametrisation. 
The second model we consider is an $f(R)$ model studied in \cite{giannantonio2010new,Hu:2013aqa} 
\bea\label{frmu}
\mu(a,k)
= \frac{1}{1-1.4\times 10^{-8}(\lambda/{\rm Mpc})^{2}a^{3}}\frac{1+\frac{4}{3}\lambda^{2}k^{2}a^{4}}{1+\lambda^{2}k^{2}a^{4}}.
\eea
The resulting equation of motion from varying the modified Einstein-Hilbert action with respect to the metric introduces a scalar degree of freedom $f_{R} = df / dR$, the \textit{scalaron}, whose Compton wavelength $\lambda$ (in Eq.~\ref{frmu}) at present can be expressed in terms of its dimensionless counterpart $B_{0}$ as $\lambda^{2} = B_{0}/(2H_{0}^{2})$. The general expression of the dimensionless Compton wavelength is given by \citep{song2007cosmological} $$B = \frac{f_{RR}}{1 + f_{R}}\frac{dR}{d{\rm ln}a}\left(\frac{d{\rm ln} H}{d{\rm ln} a}\right)^{-1}\, ,$$
where $f_{RR}$ is the second derivative of $f(R)$ with respect to the Ricci scalar $R$.
As highlighted in \cite{song2007large}, a one parameter family of $f(R)$ models labelled by $B_{0}$ exists for any given background expansion history.

In each of these two models we only have one additional parameter to constrain, $E_{11}$ or $B_0$. Current constraints on these parameters from~\cite{Ade:2015rim} are $B_0< 8.6\times 10^{-5}$ $(95 \%\> \rm CL)$ and $E_{11}=-0.30^{+0.18}_{-0.30}$ (68\% CL), obtained from a combination of Planck CMB temperature, polarisation, weak lensing, BAO, and RSD. In Fig.~\ref{dipole_plot} (left panel), we show the dipole at $z$ = 0.15 for these two models -- solid green for the scale-independent model with $E_{11} = 0.06$; and solid orange for the $f(R)$ model with $B_{0} = 0.1$ -- compared to $\Lambda$CDM in solid black. 
The shaded regions show the error bars on the dipole, calculated with the specifications of a survey like SKA phase 2 (see Section~\ref{sec:forecasts} for details). The light grey corresponds to an error on the convergence of $\sigma_{\kappa} = 0.8$, whereas the dark grey is for $\sigma_{\kappa} = 0.3$. We see that in the range $80 - 150$ Mpc/h, deviations from the GR prediction of the order of $15\%$ for the scale-independent model and $\geq 20 \%$ for the  $f(R)$ model are clearly visible in the Doppler magnification dipole.

In this figure we also compare the flat-sky approximation (dotted lines) to the full-sky calculation, finding reasonable agreement on relevant scales. In particular, the departure from the flat-sky approximation happens at the same scale in the three models, despite the different $k$-dependence of their growth rate. Note that in the following we will use the full-sky expression for the dipole, since the surveys we are interested in will cover large areas of the sky, allowing a measurement of the dipole up to large separations, where the flat-sky approximation breaks down.

In the right panel of Fig.~\ref{dipole_plot}, we plot the percentage difference between $\Lambda$CDM and the two models. In the scale-independent model, the deviation does not depend on separation $d$. As expected, the deviation decreases towards higher redshift, due to the function $\Omega_{\rm DE}(a)$ in Eq.~\eqref{debasedmu}, which suppresses deviations from GR at high redshift. The $f(R)$ model, on the other hand, has a distinct scale dependence. The function $\mu$ in Eq.~\eqref{frmu} deviates from GR at both large scales, where $\mu(a,k)\rightarrow [1-1.4\times 10^{-8}(\lambda/{\rm Mpc})^{2}a^3]^{-1}$, and small scales, where $\mu(a,k)\rightarrow 4/3[1-1.4\times 10^{-8}(\lambda/{\rm Mpc})^{2}a^3]^{-1}$. As a consequence, the dipole exhibits departure from $\Lambda$CDM at both small and large separations. The scale at which $\mu$ transitions from one asymptotic value to the other is governed by the parameter $B_0$. In particular, decreasing $B_0$ tends to shift this transition to smaller scales. This would in turn shift the deviations in the dipole to smaller scales. The $f(R)$ model also has the specificity to have a redshift dependence which depends on separation: at small separations the deviations from $\Lambda$CDM decrease with redshift, whereas at large separations they increase with redshift. To understand this behaviour, we plot in Fig.~\ref{fig:GDpercent} the relative deviations between $f(R)$ and $\Lambda$CDM in the functions $\mathcal{G}(k,z)$ and $D(z,k)/\mu$, which enter in the dipole through Eqs.~\eqref{lambda} and~\eqref{nu} and govern its redshift-dependence. We see that at large $k$ these two functions are larger in $f(R)$ than in $\Lambda$CDM. At small $k$ however, these functions are larger in $\Lambda$CDM. Looking at the amplitude of these deviations, we see that both at small $k$ and at large $k$ the amplitude decreases with redshift. This is somehow expected since $\mu$ decreases with redshift. However, since the deviations change sign, there is a transitional range in between, where the deviations increase with redshift. Once the functions are integrated over $k$ to obtain the correlation function, this transitional range seems to dominate at large separation, leading to an overall increase in the deviations with redshift.

\begin{figure}
\includegraphics[width=\columnwidth]{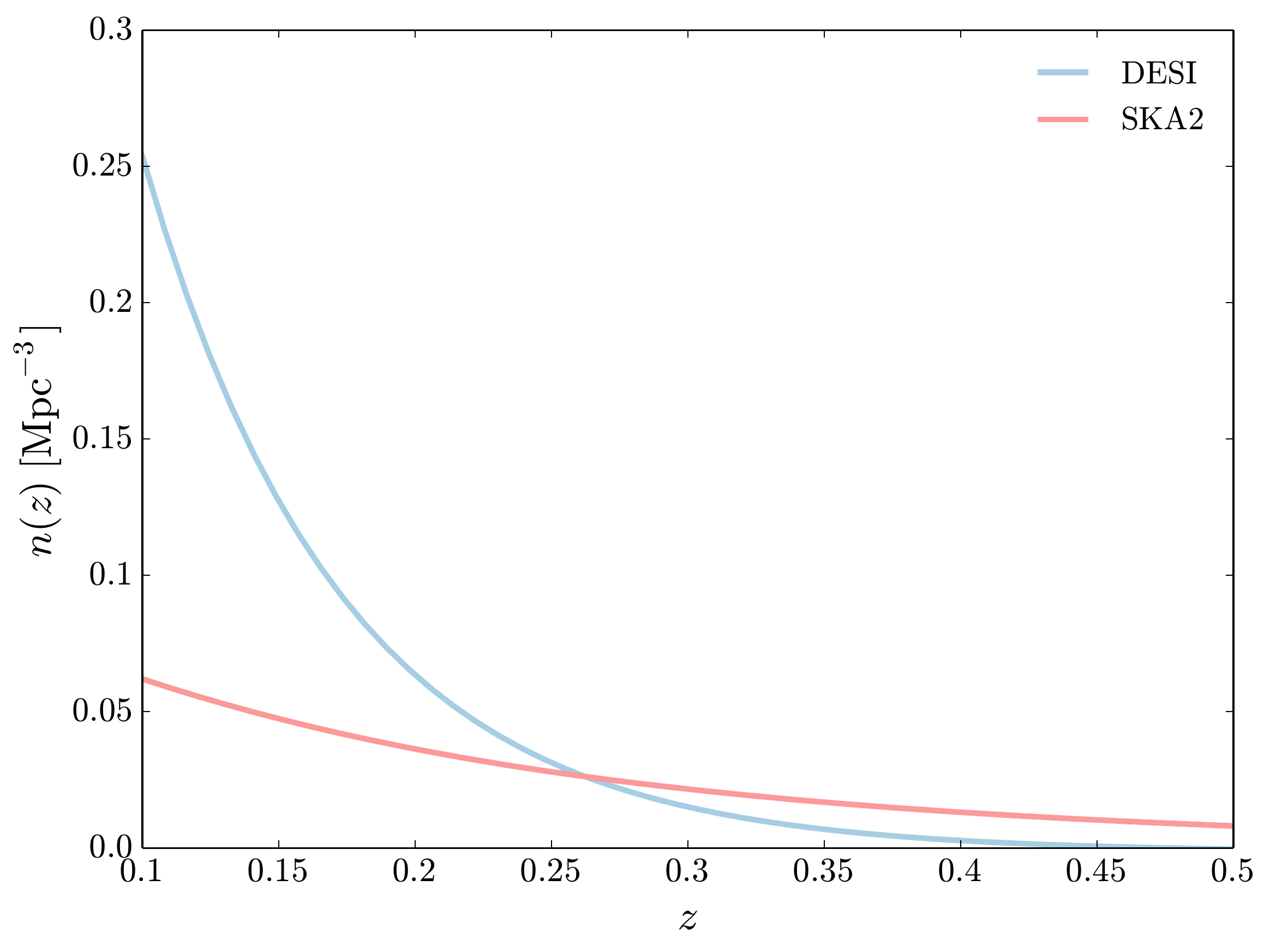}
\caption{Number density of spectroscopically-detected galaxies as a function of redshift, for DESI (red) and SKA2 (blue).}
\label{fig:dndz}
\end{figure}

For comparison, we plot in Fig.~\ref{rsd_plot} the percentage difference in the monopole and the quadrupole of redshift-space distortions (i.e. the monopole and quadrupole of $\langle \Delta \Delta\rangle$), induced by the two models. We see that, for the $f(R)$ model, the relative deviations in the monopole of RSD are significantly larger than those in the dipole. The relative deviations in the quadrupole on the other hand are quite similar to those in the dipole. This suggests that the $f(R)$ model generates larger deviations in the density than in the velocity. However, since modifications in the density are degenerate with the bias, the constraining power on $B_0$ is expected to be governed by deviations in the quadrupole.

Comparing Fig.~\ref{rsd_plot} with Fig.~\ref{dipole_plot}, we also see that the deviations in the dipole clearly increase with separation, whereas those in the monopole and quadrupole have no clear scale-dependence. This behaviour is related to the fact that Doppler magnification is directly sensitive to peculiar velocities, whereas RSD are sensitive to their gradient. As a consequence the dipole contains a factor $k/\mathcal{H}$ less than RSD, which gives more weight to larger scales. The dipole is therefore particularly well adapted to test modifications of gravity in the linear regime.

For the scale-independent model, we see that the deviations in the RSD monopole and quadrupole are of the same order of magnitude as those in the Doppler magnification dipole. This is not surprising, since in this case the functions $D$, $\mu$ and $\mathcal{G}$ can be taken out of the integrals over $k$. The different weighting in $k$ has consequently no impact on the amplitude of the deviations, leading to similar results for the three multipoles.

In the next section, we study forecasts for the overall sensitivity to the $B_0$ and $E_{11}$ parameters.

\section{Forecasts for future galaxy surveys} \label{sec:forecasts}

We now present predicted constraints on cosmological parameters in each model, using Fisher matrices, to show how deviations from GR can be constrained with the Doppler magnification dipole. 
We consider the set of parameters $h, \Omega_{m}, \Omega_{b}$ together with $E_{11}$ for the scale-independent model and $B_{0}$ for the $f(R)$ model. The fiducial values we choose are those of $\Lambda$CDM+GR with $h = 0.68,\>\Omega_{m} = 0.3028,\> \Omega_{b} = 0.048$ and the MG parameters zero. We fix the other cosmological parameters to their fiducial value: $n_s=0.96$, and $\sigma_8 = 0.83$.

\subsection{Galaxy survey specifications} \label{sec:surveys}

We assess the ability of two spectroscopic galaxy redshift surveys to constrain GR with Doppler magnification observations. The first, DESI \citep{2015DESI}, is expected to begin in 2018, and will yield multiple spectroscopic galaxy samples from a 5-year survey over a 14,000 deg$^2$ footprint. The sample of most relevance to Doppler magnification is the Bright Galaxy Sample (BGS), which covers the redshift range $0.05 \le z \lesssim 0.4$, with a median redshift of $z \simeq 0.2$. BGS galaxies will be selected from existing $r$-band imaging from DECam and Bok 90Prime, but $g$, $z$, and $3-4\mu$m band imaging will also be available \citep{2015DESI}. This provides multiple avenues for measuring the galaxy sizes to estimate $\kappa$, while DESI itself will provide high-resolution spectroscopic redshifts.

The second survey we consider is a HI galaxy survey on Phase 2 of the Square Kilometre Array (SKA), which is expected to enter operation in the late 2020s, potentially yielding a deep hemispherical ($\sim$20,000 deg$^2$) survey on the southern sky around 2030. Spectroscopic redshifts are estimated from detections of the 21cm emission line of neutral hydrogen, while sizes can be estimated from imaging of either resolved radio continuum emission, or cross-matched optical counterparts of the radio galaxies. While an SKA2 galaxy survey is expected to be sample variance limited over $0 \le z \lesssim 1.5$, we will focus on the $z \le 0.5$ range here, where the contamination from the lensing convergence is negligible (as shown in~\cite{Bonvin:2016dze}, in this regime it reaches at most 7\% at large separations $\gsim180$\,Mpc/$h$).

For both surveys, we bin the expected galaxy number density into tophat redshift bins of width $\Delta z = 0.1$, covering the range $0.1 \le z \le 0.5$. This quantity is shown in Fig.~\ref{fig:dndz}. The DESI values are taken from \cite{2016arXiv161100036D}, while the SKA2 values are taken from \cite{Bull:2015lja}. For both surveys, we fix the bias in each redshift bin to its fiducial value, given in Table~\ref{tab:bias}. Constraints for other upcoming surveys such as Euclid and LSST are broadly similar to SKA2.

\begin{table}
\caption{Fiducial value for the bias and number density $\bar n$ in $(h/{\rm Mpc})^3$ for DESI and SKA2.}
\label{tab:bias}
\begin{tabular}{ccccc}
\hline
$z$ & $b(z)$ DESI & $\bar n(z)$ DESI &$b(z)$ SKA2 & $\bar n(z)$ SKA2 \\
\hline
\hline
0.15 & 1.447 & 0.1871 & 0.623 &0.1972 \\
0.25 & 1.524& 0.0460 &0.674 & 0.1154\\
0.35 & 1.605 & 0.0098 & 0.730 & 0.0687\\
0.45 & 1.689 & 0.0010 &0.790 & 0.0417 \\
\hline
\end{tabular}
\end{table}

We use a Gaussianised Planck CMB prior in all our analyses, constructed by calculating the covariance matrix from the \cite{Ade:2015xua} MCMC chains, and then forming an effective Fisher matrix by inverting the resulting covariance matrix. This is used to constrain the standard $\Lambda$CDM parameters only; constraints on the modified gravity model parameters from the CMB are not included (although see \cite{Ade:2015rim, 2018arXiv180706209P} for Planck analyses that do include them). It also excludes CMB lensing information. The main reason for choosing to ignore information from the CMB in this way is that we wish to focus on how the Doppler magnification effect is able to {\it directly} constrain modified gravity scenarios, rather than studying its role in (e.g.) breaking degeneracies within the CMB-derived parameters to yield better constraints. Some information from the CMB is nevertheless necessary to help fix the various background parameters that would otherwise be poorly constrained by Doppler magnification alone. A more holistic analysis that includes information from contemporary surveys (such as CMB lensing and redshift-space distortions) is left for future work.

An expression for the covariance matrix was calculated in~\cite{Bonvin:2016dze}. It contains three types of contributions. First there is a contribution from cosmic variance: the cosmic variance in the number counts $\Delta$, the cosmic variance in the convergence $\kappa$, and the covariance between the two (since they trace the same underlying perturbations). Second, the covariance is affected by the shot noise in the galaxy number counts, which depends on the number density of galaxies $\bar n$. And finally it contains a contribution from the intrinsic error on the size measurement, that we denote by $\sigma_\kappa$. We obtain
\begin{align}
&{\rm cov}[\xi^{\Delta \kappa_{\vv}}](z, d, d')=\frac{9}{V}\left(1-\frac{1}{\HH r}\right)^2\left( \frac{b^2}{5}+\frac{2bf}{7}+\frac{f^2}{9}\right)f^2\nonumber\\
&\hspace{1cm}\times \frac{\HH^2}{\pi^2}\int dk P^2_{\delta\delta}(k,z)j_1(kd)j_1(kd')\nonumber\\
&+ \frac{9}{2}\sigma_\kappa^2\frac{\ell_p^3}{V}
\left(\frac{b^2}{3}+\frac{2bf}{5}+\frac{f^2}{7}\right)
\frac{1}{\pi^2}\int dk k^2 P_{\delta\delta}(k,z)j_1(kd)j_1(kd')\nonumber\\
&+\frac{3}{4\pi}\frac{\sigma_\kappa^2}{\bar n V}\left(\frac{\ell_p}{d}\right)^2\delta_K(d-d')\, , 
\label{eq:cov}
\end{align}
where $P_{\delta\delta}(k,z)$ is the matter power spectrum at redshift $z$, $\ell_p$ denotes the size of the cubic pixel in which we measure $\Delta$ and $V$ is the volume of the survey (or of the redshift bin of interest). Note that in the calculation of the covariance we account for correlations between different pixel's separations, but we neglect correlations between different redshift bins. Since the size of the bins that we use is relatively large, this is a good approximation. 

\begin{figure*}
\includegraphics[width=\columnwidth]{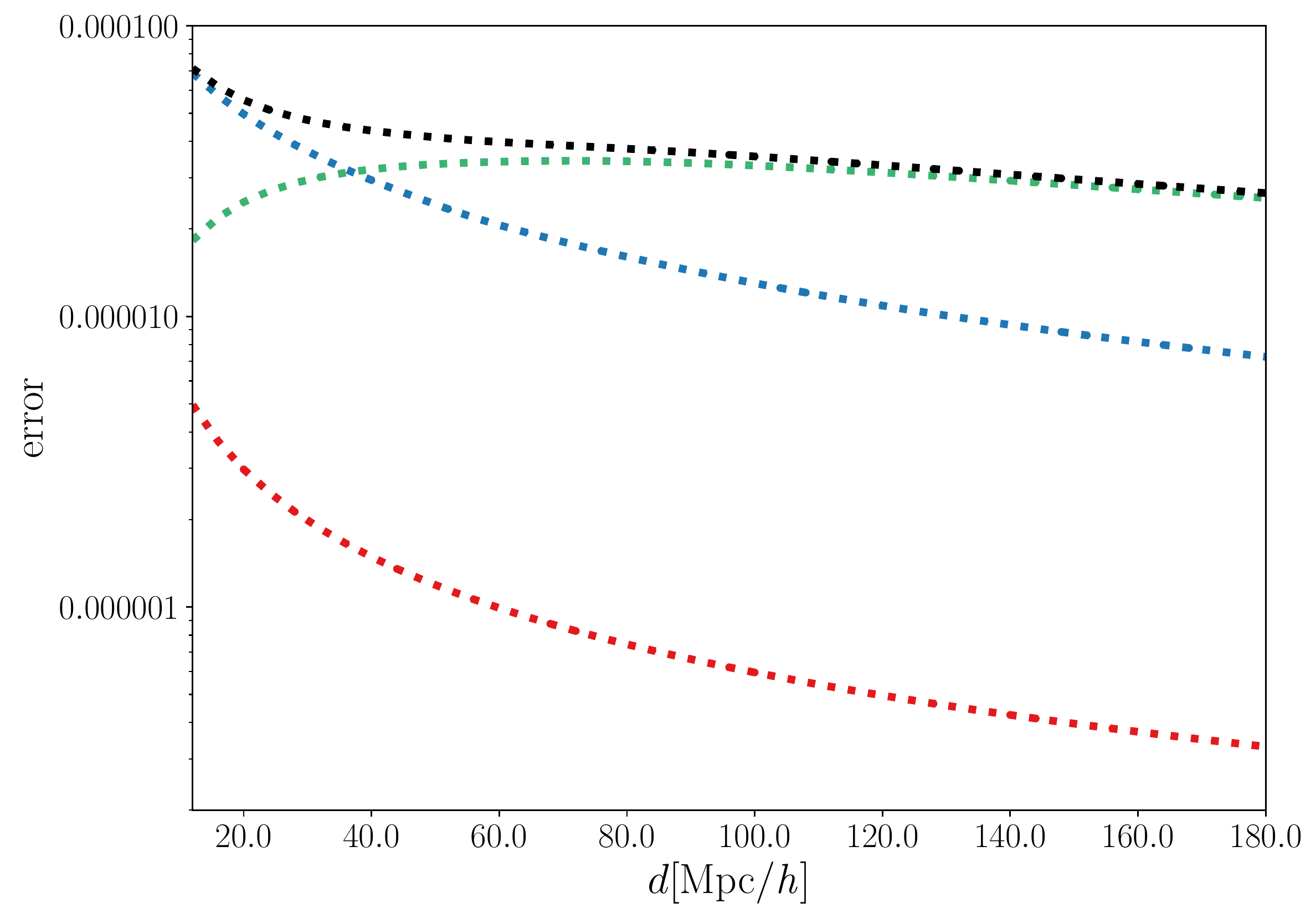}
\includegraphics[width=\columnwidth]{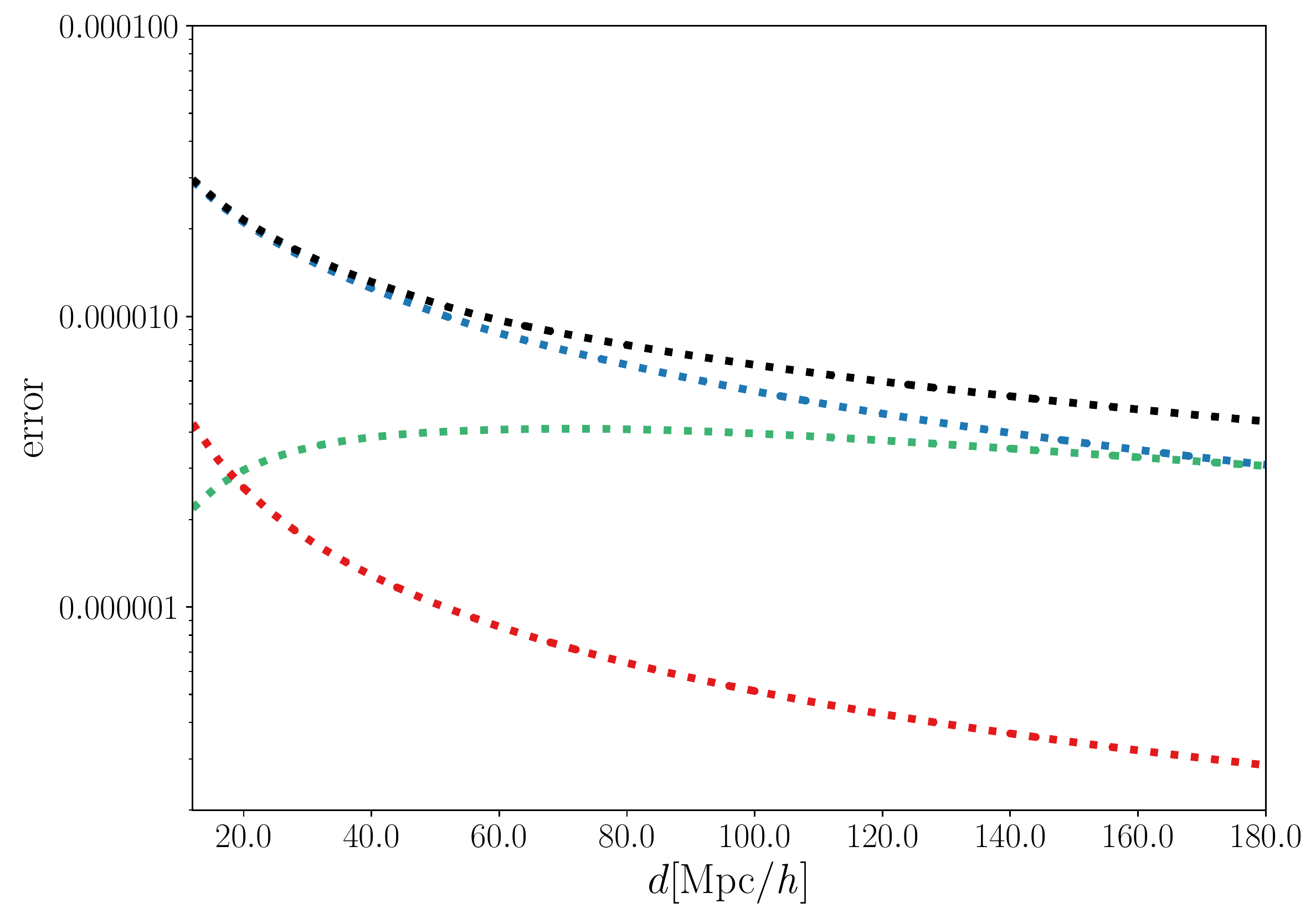}
\caption{Error on the dipole, calculated from Eq.~\eqref{eq:cov}, using the specifications of SKA2 and $\sigma_\kappa=0.3$. The left panel shows the error at $z=0.15$ and the right panel at $z=0.45$. The pixel size is $\ell_p=4$\,Mpc/$h$. The green dots show the pure cosmic variance (first two lines of Eq.~\eqref{eq:cov}), the blue dots show the product of size measurement error $\sigma_\kappa$ and cosmic variance of $\Delta$ (third line of Eq.~\eqref{eq:cov}), the red dots show the product of size measurement error and shot noise (fourth line of Eq.~\eqref{eq:cov}), and the black dots show the total.}\label{cov_plot}
\end{figure*}

In the following we choose two representative values for $\sigma_{\kappa} = 0.3,\> 0.8$. We refer the interested reader to \cite{alsing2015weak, Bonvin:2016dze} for a discussion on how the value of $\sigma_{\kappa}$ may change depending on the type of galaxies in a given survey. In Fig.~\ref{cov_plot} we show the different contributions to the error, as a function of separation (i.e.\ $\sqrt{{\rm cov}(d,d)}$), for two redshift bins of width 0.1 centered around $z=0.15$ (left panel) and $z=0.45$ (right panel). The green dots are the contribution from the first two lines of Eq.~\eqref{eq:cov}, which are due to the cosmic variance of $\Delta$ and of $\kappa$. The blue dots are the contribution from the third line of Eq.~\eqref{eq:cov}, due to the product of size measurement error $\sigma_\kappa$ and cosmic variance of $\Delta$. The red dots are the contribution from the last line of Eq.~\eqref{eq:cov}, due to the product of size measurement error and shot noise. The black dots show the total. Comparing the two panels, we see that the terms involving cosmic variance are significantly smaller at $z=0.45$ than at $z=0.15$ due to the larger volume covered at higher redshift. The red dots are similar in the two panels, since shot noise is sensitive to the total number of galaxies in the redshift bin, which is very similar in the two bins displayed here: the larger volume at $z=0.45$ is compensated by a smaller number density. Note that increasing $\sigma_\kappa$ enhances the blue and red contributions, with respect to the green one. In~\cite{Bonvin:2016dze}, we neglected the pure cosmic variance contribution (green dots), which is a reasonable approximation for $z\geq 0.25$, but not for the lowest redshift bin. We checked however that this does not change the overall signal-to-noise of the dipole.

\begin{figure}
\includegraphics[width=\columnwidth]{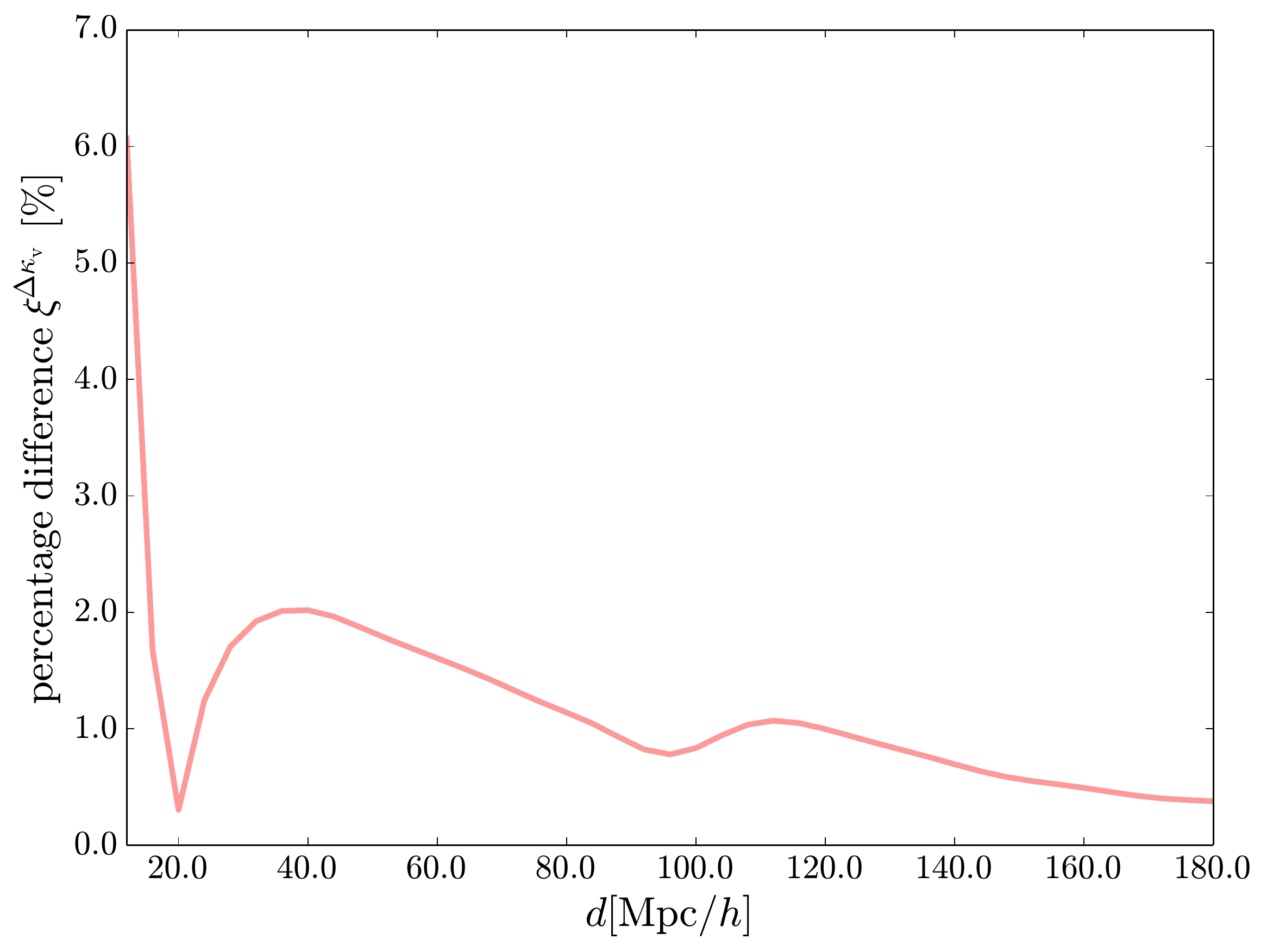}
\caption{Percentage difference between the linear dipole and a HaloFit-based approximation of the non-linear dipole (see footnote~\ref{foot:non-linear}) in $\Lambda{\rm CDM}$.}
\label{compare_l_nl_dipole}
\end{figure}

\subsection{Results} \label{sec:results}

In this section, we present forecasts for how well Doppler dipole measurements with DESI and SKA2 galaxy survey will be able to constrain the scale-independent and $f(R)$ modified gravity parametrisations discussed above.

Since our expression for the dipole is based on linear perturbation theory, we restrict our analysis to separations $d$ = 40-180 Mpc/$h$. In Fig.~\ref{compare_l_nl_dipole} we estimate the impact of non-linearities on the dipole and we show that $d\geq 40$\,Mpc/$h$ is a conservative minimum separation, for which the effect of non-linearities is less than 2$\%$.\footnote{To calculate the impact of non-linearities on the dipole we use the following approximation: we use the linear continuity equation to relate the velocity to the density, and then we calculate the non-linear density power spectrum with HaloFit. This procedure is not correct, since in the non-linear regime the continuity equation is modified. However, it allows us to evaluate at which scales non-linearities become relevant. \label{foot:non-linear}}

\begin{figure*}
	\includegraphics[width=\columnwidth]{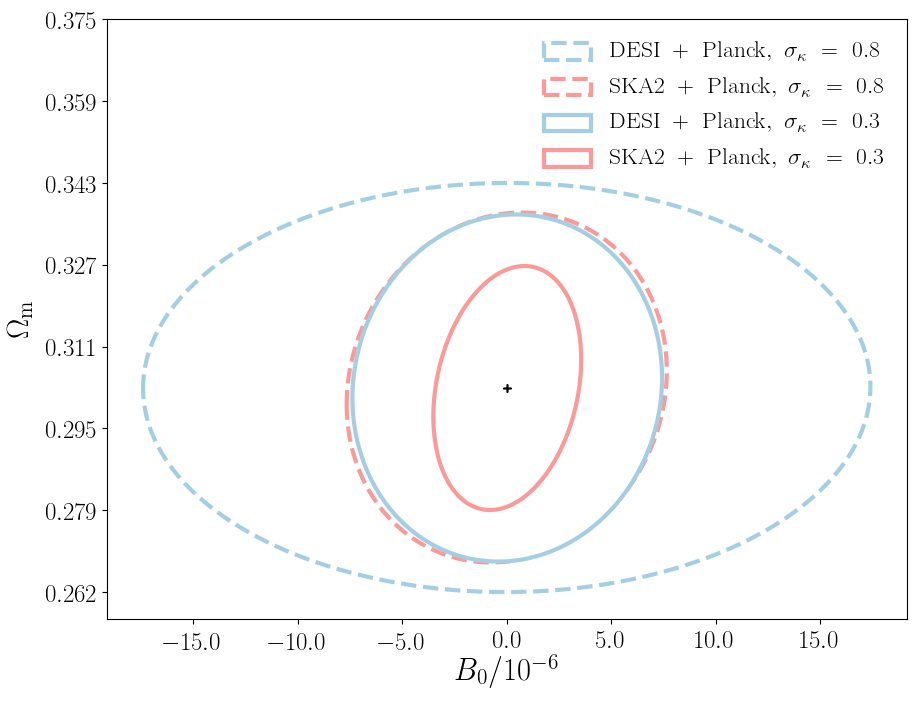}
	\includegraphics[width=\columnwidth]{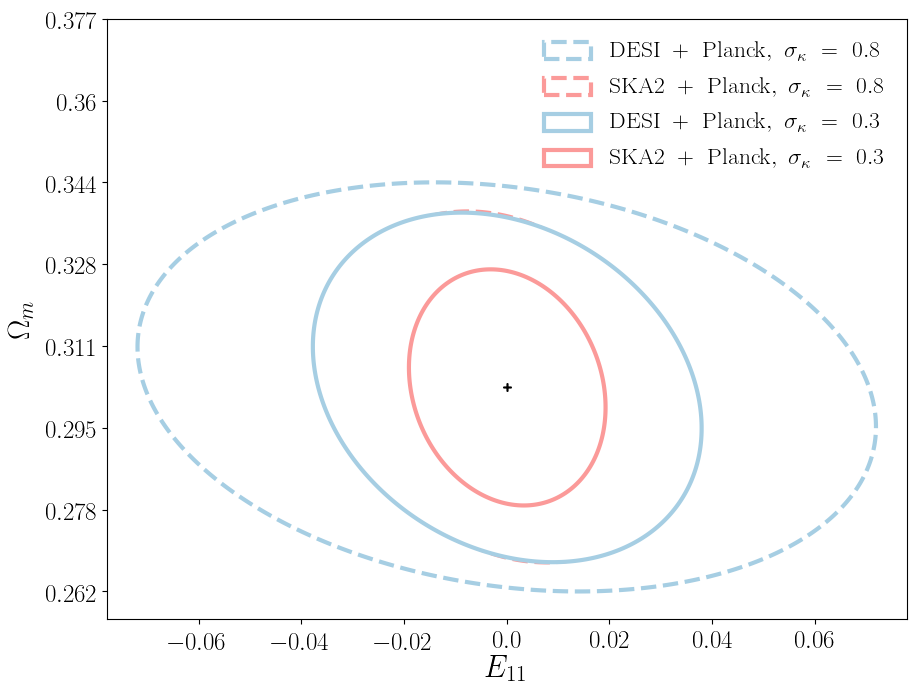}
    \caption{Joint marginalised constraints $B_{0}-\Omega_{m}$ for the $f(R)$ model (left) and $E_{11}-\Omega_{m}$ for the scale-independent model (right). Dashed blue and solid  blue ellipses are $68\%$ CL for the DESI survey, considering $\sigma_{\kappa} = 0.8$ and $\sigma_{\kappa} = 0.3$ respectively. Dashed red and solid red ellipses are $68\%$ CL for the SKA2 survey, using $\sigma_{\kappa} = 0.8$ and $\sigma_{\kappa} = 0.3$ respectively.}    \label{B0Om}
\end{figure*}

The left panel of Fig.~\ref{B0Om} shows the joint constraints in the $f(R)$ model on $\Omega_{m}$ and the parameter $B_{0}$, marginalised over the other parameters. The constraints are sensitive to the value of the error on the size measurement: the constraints for $\sigma_{\kappa} = 0.3$ are better than those corresponding to $\sigma_{\kappa} = 0.8$ by a factor of $\sim 2$. It is worth pointing out that, without a Planck prior, the value of $\sigma_\kappa$ predominantly affects the diagonal of the forecast parameter covariance matrix, while it can affect both the resulting constraint and the correlation between parameters when a prior is included. The marginalised constraints on $B_{0}$ ($95 \%\> \rm CL $), obtained by combining the Doppler magnification dipole with Planck, are $B_{0}< 1.2\times 10^{-5}$ with DESI and $B_{0}< 5.7\times 10^{-6}$ with SKA2, assuming $\sigma_{\kappa}$ = 0.3. Including scales down to $d$ = 20 Mpc/$h$ tightens the constraints by one order of magnitude: $B_{0}< 1.0\times 10^{-6}$ with DESI and $B_{0}< 5.1\times 10^{-7}$ with SKA2. This shows that the constraining power of the dipole is not too strongly degraded by limiting the analysis to linear scales.

For comparison, the current constraints on $B_{0}$ from \cite{Ade:2015rim} are $B_0< 8.6\times 10^{-5}$ $(95 \%\> \rm CL)$, obtained from a combination of Planck CMB temperature, polarisation, weak lensing, BAO, and RSD. The Doppler magnification dipole is therefore expected to improve the current constraints by one order of magnitude with SKA2.

In the right panel of Fig.~\ref{B0Om}, we show the joint constraints for the scale-independent model on $\Omega_{m}$ and the parameter $E_{11}$, marginalised over the other parameters. The marginalised constraints on $E_{11}$ ($68 \%\> \rm CL $), obtained by combining the Doppler magnification dipole with Planck, are $E_{11}< 0.06$ with DESI and $E_{11}< 0.03$ with SKA2, assuming $\sigma_{\kappa}$ = 0.3. Comparing with current constraints from \cite{Ade:2015rim}: $E_{11}=-0.30^{+0.18}_{-0.30}$ (68\% CL), we see that the Doppler magnification dipole is again expected to improve the constraints by one order of magnitude.

To understand the different constraining power of the Doppler magnification dipole versus RSD, it is first informative to compare the cumulative signal-to-noise of these two probes. The signal-to-noise for the Doppler magnification dipole within SKA2, for $0.1\leq z\leq 0.5$ and $40 \leq d\leq 180\,$Mpc/$h$ is of 70 for $\sigma_\kappa=0.3$. For RSD, we can calculate the signal-to-noise of the monopole and quadrupole, using the specifications of the CMASS DR11 sample~\citep{Samushia:2013yga} used for the Planck constraints~\citep{Ade:2015rim}. Using the publicly available code COFFE~\citep{tansella2018coffe} to calculate the signal and covariance matrices, we obtain a cumulative signal-to-noise of 80 for $24\leq d\leq 152\,$Mpc/$h$. This shows that the precision with which the Doppler magnification dipole will be measured with SKA2 is similar to the precision of current RSD measurements. 

Since the deviations in the Doppler magnification dipole are similar to those in the multipoles of RSD (see Figs.~\ref{dipole_plot} and~\ref{rsd_plot}), we would then expect similar constraints on $B_0$ and $E_{11}$ from these two probes. The fact that we find instead an order of magnitude improvement  with the Doppler magnification dipole is due to the fact that in our analysis we fix the value of the bias. This automatically breaks the degeneracy between modifications of gravity and bias evolution. On the other hand, RSD analyses consider the bias as a free parameter. Combined measurements of the monopole and quadrupole allow us then to measure separately the combination $b(z)D(z,k)/\mu(z,k)$ and $f(z,k)D(z,k)/\mu(z,k)$, and therefore to break the degeneracy between bias evolution and modifications of gravity for a given model. However, due to measurement uncertainty, this degeneracy is in practice only partly broken, leading to a degradation of the constraints compared to our analysis. Our analysis is therefore in this sense too optimistic: including the bias as a free parameter and marginalising over it would degrade the constraints from the Doppler magnification dipole. A full analysis should provide joint constraints from the monopole, quadrupole and hexadecapole of RSD and the Doppler magnification dipole. This would however require to compute the covariance of the dipole with the RSD multipoles, which is highly non-trivial and beyond the scope of this paper. For this reason we assume that the bias will be tightly determined by the RSD multipoles, and we use the Doppler magnification dipole as an additional probe of the growth rate.

\begin{figure*}
\includegraphics[width=1.\columnwidth]{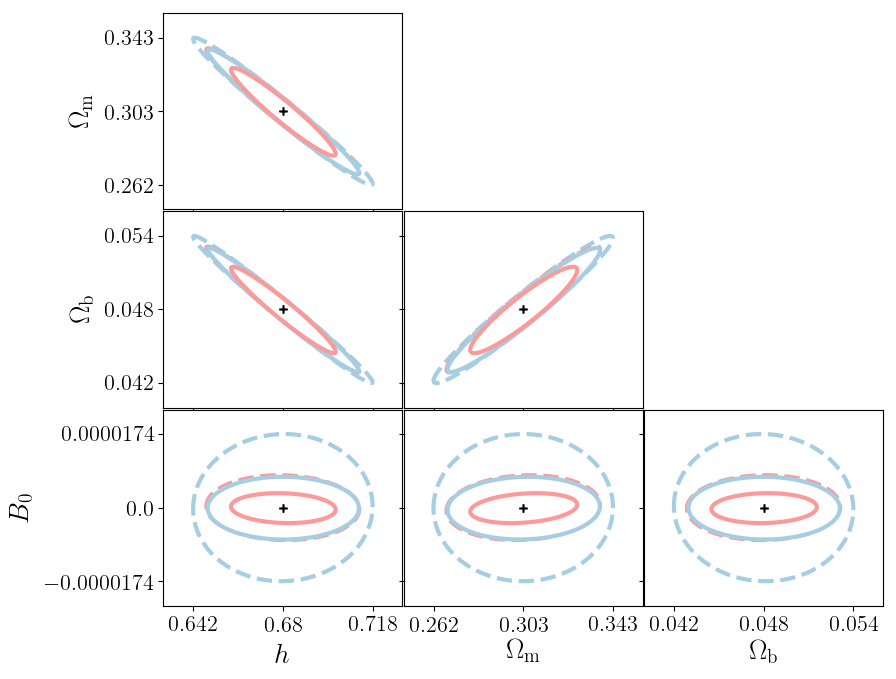}
	\includegraphics[width=1.\columnwidth]{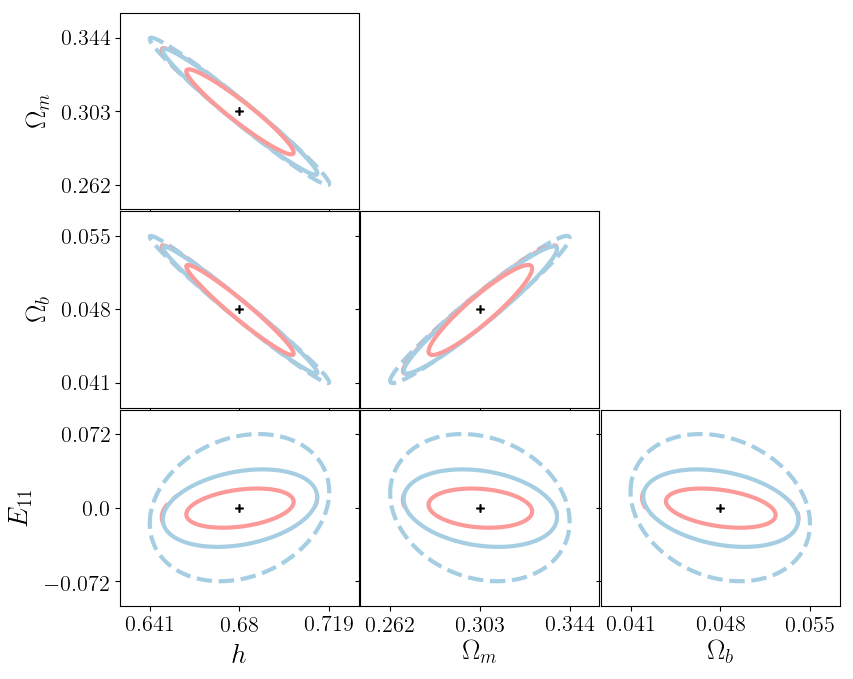}
    \caption{Constraints on all the parameters in the $f(R)$ model (left) and the scale-independent model (right). All the ellipses are $68\%$ CL. Dashed blue corresponds to DESI with $\sigma_{\kappa} = 0.8$, solid blue to DESI with $\sigma_{\kappa} = 0.3$, dashed red corresponds to SKA2 with $\sigma_{\kappa} = 0.8$ and solid blue to SKA with $\sigma_{\kappa} = 0.3$.}
    \label{all_fR_DE}
\end{figure*}

Finally, let us mention that the Planck constraints on $B_0$ are highly sensitive to how a degeneracy between $\tau$, $A_s$, and $B_0$ is broken. With Planck CMB measurements alone, the upper limit is $B_0 < 0.79$ (95\% CL), which is reduced to $< 0.69$ when BAO, Type Ia supernovae, and $H_0$ (`BSH') measurements are added. Further adding RSD measurements reduces the upper limit to $<0.90 \times 10^{-4}$ however -- an improvement of around four orders of magnitude! The explanation for this dramatic improvement is that even relatively weak constraints on structure formation (i.e. those providing measurements of $\sigma_8(z)$) are sufficient to break the degeneracy with $\tau$ and $A_s$ and therefore tightly constrain $B_0$. This effect is not captured by our forecasts, which only use the Planck constraints as a prior on the standard cosmological parameters; if $B_0$ were included in our Planck prior Fisher matrix, a similar effect would be observed when combined with the Doppler dipole Fisher matrix, as this also constrains the growth rate of structure. 

For completeness, forecast constraints on all parameters of the scale-independent and $f(R)$ models that we considered are shown in the left and right panels of Fig.~\ref{all_fR_DE} respectively.

The marginalised constraints on the $B_0$ and $E_{11}$ parameters are shown as a function of redshift in Tables~\ref{constraints_per_bin_fR} and~\ref{constraints_per_bin_DE} respectively. For both models, and both SKA and DESI, the constraints are best in the lowest redshift bin, and get gradually worse with increasing redshift. This behaviour is more pronounced for the constraints on $B_0$ than on $E_1$. From Fig.~\ref{cov_plot} we see that the variance decreases with redshift, due to the larger volume available, which reduces the cosmic variance. On the other hand, the signal itself also decreases with redshift due to the coefficient $1/(\HH r)-1$  in front of the dipole. The balance between these two effects is furthermore weighted by the strength of the deviations from $\Lambda$CDM, which is different for the two models. In both cases, the regime $0.1\leq z\leq 0.5$ is well adapted to the analysis. This regime has also the advantage that the gravitational lensing contribution is always strongly suppressed and can therefore be safely neglected. It is notable that the predicted constraints from SKA2 are always around a factor of two better than for DESI, irrespective of the value of $\sigma_\kappa$ that is assumed. This is primarily due to the significantly larger survey area of SKA2, despite it having a lower number density than the DESI sample in the lowest redshift bins.

\begin{figure*}
    \includegraphics[width=\columnwidth]{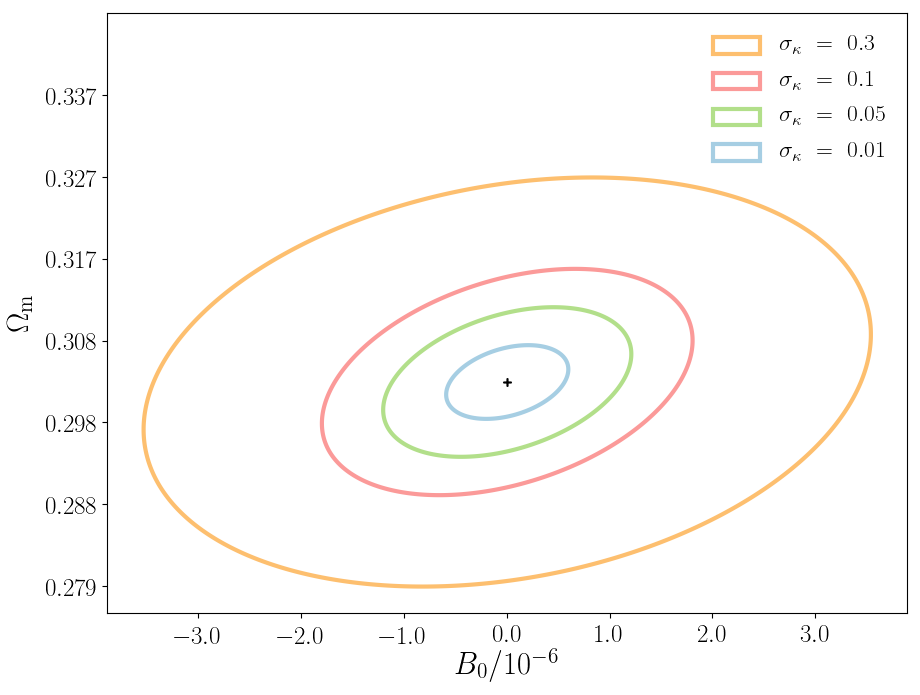}
    \includegraphics[width=\columnwidth]{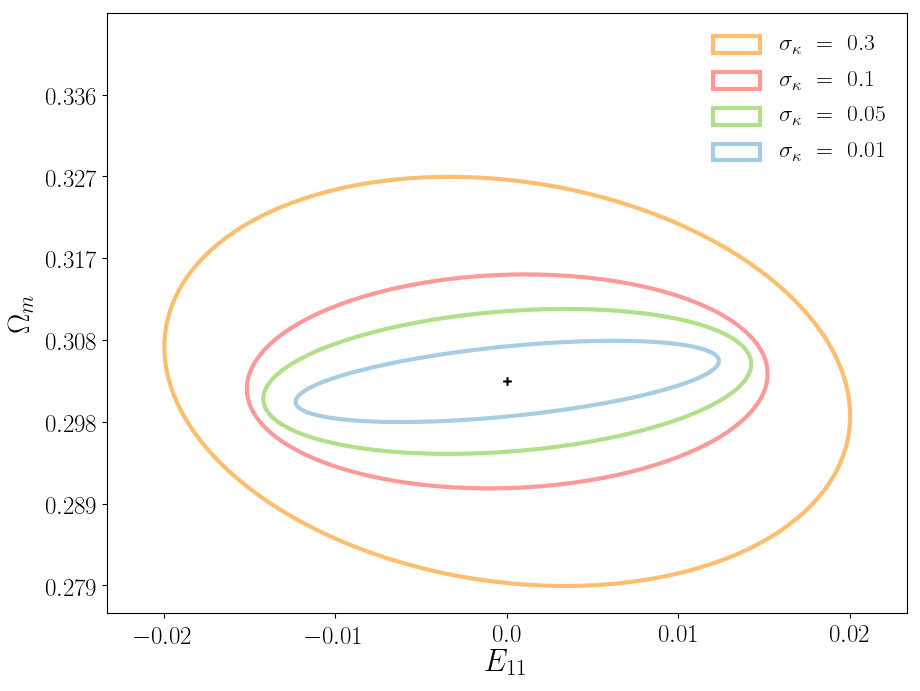}
    \caption{Constraints on $\Omega_{\rm m}$ and the $B_0$ parameter of the $f(R)$ model (left) and $E_{11}$ parameter of the scale-independent model (right) for SKA2 + Planck ($68\%$ CL), for several different values of $\sigma_\kappa$.}
    \label{fig:sigmakappa}
\end{figure*}

In all of the cases we have considered so far, the Doppler dipole measurement is strongly affected by the intrinsic error on the galaxy size, $\sigma_\kappa$. This error is especially important at larger $z$, where the cosmic variance is small and the intrinsic error becomes therefore the dominant source of error (see Fig.~\ref{cov_plot}). Fig.~\ref{fig:sigmakappa} shows results for several different values of $\sigma_\kappa$, from the more optimistic value of $0.3$ that was used in the forecasts above, to a highly optimistic value of $0.01$. The latter value is quite extreme, and we make no claim that it can be achieved in practice -- galaxies are complex objects formed by messy nonlinear processes, and so there will always be a significant amount of scatter in the size distribution of any population \cite[c.f.][]{alsing2015weak}. There is some hope that galaxy samples (or proxy observables) can be selected to reduce the scatter however; for example, the relationship between HI mass and disk radius has a particularly low scatter of $\approx 0.06$ dex ($\sigma_\kappa \approx 0.14$) \citep{2016MNRAS.460.2143W}. As shown in Fig.~\ref{fig:sigmakappa}, the gain from reducing $\sigma_\kappa$ from 0.3 to 0.1 is a factor of $\sim\! 2$ for the $f(R)$ model with $B_{0} < 2.94\times 10^{-6}$, but only about 26$\%$ for the scale-independent model with $E_{11} < 2.3\times 10^{-2}$ ($95\%$ CL).
The prospects for reducing $\sigma_\kappa$ even further are unclear, as this would likely require a galaxy size measurement method to be devised that does not inherently depend on statistical galaxy size distributions, e.g. by using some kind of non-statistical standard ruler. We are not currently aware of any such method that could be used to make practical measurements however.

\begin{table}
 \caption{Marginalised constraints on the $B_{0}$ parameter, obtained at each redshift bin with two different values of $\sigma_{\kappa}$.}
 \label{constraints_per_bin_fR}
 
 \begin{tabular}{llll}
  \hline
  $\sigma_{\kappa}$ & $z$ & SKA2 ($95\%$ CL) & DESI ($95\%$ CL) \\
  \hline
  \hline
  \multirow{4}{4em}{$0.3$}
  & 0.15 & $< 7.15\times 10^{-6}$ & $< 1.55\times 10^{-5}$\\[2pt]
  & 0.25 & $< 1.04\times 10^{-5}$& $< 2.32\times 10^{-5}$\\[2pt]
  & 0.35 & $< 1.60\times 10^{-5}$ & $< 3.56\times 10^{-5}$\\[2pt]
  & 0.45 & $< 2.49\times 10^{-5}$ & $< 5.61\times 10^{-5}$ \\[2pt]
  \hline
  \multirow{4}{4em}{$0.8$}
  & 0.15 & $< 1.56\times 10^{-5}$& $< 3.60\times 10^{-5}$ \\[2pt]
  & 0.25 & $< 2.48\times 10^{-5}$& $< 5.72\times 10^{-5}$\\[2pt]
  & 0.35 & $< 3.95\times 10^{-5}$& $< 9.03\times 10^{-5}$\\[2pt]
  & 0.45 & $< 6.25\times 10^{-5}$ & $< 1.44\times 10^{-4}$ \\[2pt]

  \hline
 \end{tabular}
 
\end{table}

\section{Conclusions} \label{sec:conclusions}

We have shown the potential of using the Doppler magnification dipole, prescribed by \cite{Bonvin:2016dze}, to constrain departures from GR. 
To illustrate the sensitivity of Doppler magnification to modified gravity, we have chosen two toy models in the Parametrised Post-Friedmann formalism, one with a scale-independent growth rate, and one $f(R)$ model with a scale-dependent growth rate.

In the quasi-static regime within the scales of interest, we have derived an expression for the peculiar velocity in the two models. With our choice of parametrisation, the peculiar velocity is sensitive to the function $\mu$, which encodes deviations in the growth equation~\eqref{bardnmg}.

We have then derived the cross-correlation between the convergence and the galaxy number counts $\xi^{\Delta \kappa_{v}}$ in the two models, and we have compared it with that of GR.
As expected, the difference between the scale-independent model and GR is constant at all separations $d$. It is however redshift dependent, decreasing at higher redshift. On the other hand, the scale-dependence of the $f(R)$ model modifies the shape of the dipole. The departure from GR exhibits then a minimum at small separation and then constantly increases towards large separation. 

Since the Doppler magnification dipole should be detected with a high signal-to-noise in SKA2 and DESI, we have used these surveys to forecast constraints on the parameters $E_{11}$ and $B_0$ that encode deviations from GR in the two models. We have found that for DESI the constraints on $B_0$ and $E_{11}$ are expected to be similar to current RSD constraints. The constraints from SKA2 are expected to be one order of magnitude better. This improvement is however mainly due to the fact that we fix the bias in our analysis.

\begin{table}
 \caption{Marginalised constraints on the $E_{11}$ parameter, obtained at each redshift bin with two different values of $\sigma_{\kappa}$.}
 \label{constraints_per_bin_DE}
 \begin{tabular}{llll}
  \hline
  $\sigma_{\kappa}$ & $z$ & SKA2 ($95\%$ CL) & DESI ($95\%$ CL) \\
  \hline
  \hline
  \multirow{4}{4em}{$0.3$}
  & 0.15 & $< 5.61\times 10^{-2}$ & $< 1.19\times 10^{-1}$\\[2pt]
  & 0.25 & $< 5.08\times 10^{-2}$& $< 1.08\times 10^{-1}$\\[2pt]
  & 0.35 & $< 5.42\times 10^{-2}$ & $< 1.13\times 10^{-1}$\\[2pt]
  & 0.45 & $< 6.50\times 10^{-2}$ & $< 1.33\times 10^{-1}$ \\[2pt]
  \hline
  \multirow{4}{4em}{$0.8$}
  & 0.15 & $< 9.36\times 10^{-2}$& $< 2.10\times 10^{-1}$ \\[2pt]
  & 0.25 & $< 9.52\times 10^{-2}$& $< 2.11\times 10^{-1}$\\[2pt]
  & 0.35 & $< 1.07\times 10^{-1}$& $< 2.34\times 10^{-1}$\\[2pt]
  & 0.45 & $< 1.30\times 10^{-1}$ & $< 2.83\times 10^{-1}$ \\[2pt]
  \hline
 \end{tabular}
\end{table}

We have used four tomographic bins to get the constraints on both $B_{0}$ and $E_{11}$. To investigate which tomographic bin provides the constraining power, we have computed constraints as a function of redshift bin for the two surveys and found that the resulting constraints decrease with redshift, in other words the constraining power mainly comes from the bin at low redshift ($z = 0.15$). Overall, constraints from SKA2 are approximately twice as tight as those from DESI at all redshift bins.

To get an idea of how sensitive to the errors on size measurement the constraints are, we have chosen optimistic and pessimistic cases with $\sigma_{\kappa} = 0.3,\>0.8$ respectively. We have found that decreasing $\sigma_\kappa$ from 0.8 to 0.3 improves the constraints by a factor of 2. We have also explored how the constraints vary if we decrease $\sigma_\kappa$ from 0.3 to the (unrealistic) value of 0.01, finding an improvement by a factor of $\sim 10$ for the $f(R)$ model and by a factor of $\sim 1.6$ for the scale-independent model. This shows that the error on size measurement is the dominant source of uncertainty in the dipole.

Finally, let us mention that in our analysis we have considered only two specific models: an $f(R)$ model which modifies the growth rate at both small and large scales, and a scale-independent model. If on the other hand, we would have modifications of gravity that are significant only at large scales, then we would expect the Doppler magnification dipole to be more sensitive to these modifications than RSDs. As discussed above, the Doppler magnification dipole has one factor of $k/\mathcal{H}$ less than RSD, making it especially sensitive to modifications at large scales.

We conclude that the Doppler magnification dipole, considering future surveys like $\rm SKA$, has good prospects for investigating modification of gravity on sub-horizon scales. In the event that RSD measures a departure from GR in the future, it will be crucial to check this result with an independent probe. Our analysis shows that the Doppler magnification dipole does provide an alternative way of testing GR with peculiar velocities: despite having a lower signal to noise than RSD, it is sensitive to different systematics and is therefore complementary to RSD.

\balance
\section*{Acknowledgements}

We are grateful to N.~Maddox for useful comments. SA is funded by South African Radio Astronomy Observatory (SARAO). CB acknowledges funding by the Swiss National Science Foundation. CC was supported by STFC Consolidated Grant ST/P000592/1. DB and RM are supported by STFC Consolidated Grant ST/N000668/1. RM is also supported by South African Radio Astronomy Observatory (SARAO) and the National Research Foundation (Grant No. 75415).




\bibliographystyle{mnras}
\bibliography{dipole} 

\begin{thebibliography}{}
\makeatletter
\relax
\def\mn@urlcharsother{\let\do\@makeother \do\$\do\&\do\#\do\^\do\_\do\%\do\~}
\def\mn@doi{\begingroup\mn@urlcharsother \@ifnextchar [ {\mn@doi@}
  {\mn@doi@[]}}
\def\mn@doi@[#1]#2{\def\@tempa{#1}\ifx\@tempa\@empty \href
  {http://dx.doi.org/#2} {doi:#2}\else \href {http://dx.doi.org/#2} {#1}\fi
  \endgroup}
\def\mn@eprint#1#2{\mn@eprint@#1:#2::\@nil}
\def\mn@eprint@arXiv#1{\href {http://arxiv.org/abs/#1} {{\tt arXiv:#1}}}
\def\mn@eprint@dblp#1{\href {http://dblp.uni-trier.de/rec/bibtex/#1.xml}
  {dblp:#1}}
\def\mn@eprint@#1:#2:#3:#4\@nil{\def\@tempa {#1}\def\@tempb {#2}\def\@tempc
  {#3}\ifx \@tempc \@empty \let \@tempc \@tempb \let \@tempb \@tempa \fi \ifx
  \@tempb \@empty \def\@tempb {arXiv}\fi \@ifundefined
  {mn@eprint@\@tempb}{\@tempb:\@tempc}{\expandafter \expandafter \csname
  mn@eprint@\@tempb\endcsname \expandafter{\@tempc}}}

\bibitem[\protect\citeauthoryear{Abbott et~al.}{Abbott
  et~al.}{2016a}]{Abbott:2016blz}
Abbott B.~P.,  et~al., 2016a, \mn@doi [Phys. Rev. Lett.]
  {10.1103/PhysRevLett.116.061102}, 116, 061102

\bibitem[\protect\citeauthoryear{Abbott et~al.}{Abbott
  et~al.}{2016b}]{TheLIGOScientific:2016src}
Abbott B.~P.,  et~al., 2016b, \mn@doi [Phys. Rev. Lett.]
  {10.1103/PhysRevLett.116.221101}, 116, 221101

\bibitem[\protect\citeauthoryear{Abbott et~al.}{Abbott
  et~al.}{2017a}]{Abbott:2017vtc}
Abbott B.~P.,  et~al., 2017a, \mn@doi [Phys. Rev. Lett.]
  {10.1103/PhysRevLett.118.221101}, 118, 221101

\bibitem[\protect\citeauthoryear{Abbott et~al.}{Abbott
  et~al.}{2017b}]{Monitor:2017mdv}
Abbott B.~P.,  et~al., 2017b, \mn@doi [Astrophys. J.]
  {10.3847/2041-8213/aa920c}, 848, L13

\bibitem[\protect\citeauthoryear{Alonso, Louis, Bull  \& Ferreira}{Alonso
  et~al.}{2016}]{Alonso:2016jpy}
Alonso D.,  Louis T.,  Bull P.,   Ferreira P.~G.,  2016, \mn@doi [Phys. Rev.]
  {10.1103/PhysRevD.94.043522}, D94, 043522

\bibitem[\protect\citeauthoryear{Alsing, Kirk, Heavens  \& Jaffe}{Alsing
  et~al.}{2015a}]{Alsing:2014fya}
Alsing J.,  Kirk D.,  Heavens A.,   Jaffe A.,  2015a, \mn@doi [Mon. Not. Roy.
  Astron. Soc.] {10.1093/mnras/stv1249}, 452, 1202

\bibitem[\protect\citeauthoryear{Alsing, Kirk, Heavens  \& Jaffe}{Alsing
  et~al.}{2015b}]{alsing2015weak}
Alsing J.,  Kirk D.,  Heavens A.,   Jaffe A.~H.,  2015b, Monthly Notices of the
  Royal Astronomical Society, 452, 1202

\bibitem[\protect\citeauthoryear{Amendola et~al.}{Amendola
  et~al.}{2013a}]{Amendola:2012ys}
Amendola L.,  et~al., 2013a, \mn@doi [Living Rev. Rel.] {10.12942/lrr-2013-6},
  16, 6

\bibitem[\protect\citeauthoryear{Amendola, Kunz, Motta, Saltas  \&
  Sawicki}{Amendola et~al.}{2013b}]{Amendola:2012ky}
Amendola L.,  Kunz M.,  Motta M.,  Saltas I.~D.,   Sawicki I.,  2013b, \mn@doi
  [Phys. Rev.] {10.1103/PhysRevD.87.023501}, D87, 023501

\bibitem[\protect\citeauthoryear{Bacon, Andrianomena, Clarkson, Bolejko  \&
  Maartens}{Bacon et~al.}{2014}]{Bacon:2014uja}
Bacon D.~J.,  Andrianomena S.,  Clarkson C.,  Bolejko K.,   Maartens R.,  2014,
  \mn@doi [Mon. Not. Roy. Astron. Soc.] {10.1093/mnras/stu1270}, 443, 1900

\bibitem[\protect\citeauthoryear{Baker, Ferreira  \& Skordis}{Baker
  et~al.}{2013}]{Baker:2012zs}
Baker T.,  Ferreira P.~G.,   Skordis C.,  2013, \mn@doi [Phys. Rev.]
  {10.1103/PhysRevD.87.024015}, D87, 024015

\bibitem[\protect\citeauthoryear{Baker, Psaltis  \& Skordis}{Baker
  et~al.}{2015}]{Baker:2014zba}
Baker T.,  Psaltis D.,   Skordis C.,  2015, \mn@doi [Astrophys. J.]
  {10.1088/0004-637X/802/1/63}, 802, 63

\bibitem[\protect\citeauthoryear{{Baldi}, {Villaescusa-Navarro}, {Viel},
  {Puchwein}, {Springel}  \& {Moscardini}}{{Baldi}
  et~al.}{2014}]{2014MNRAS.440...75B}
{Baldi} M.,  {Villaescusa-Navarro} F.,  {Viel} M.,  {Puchwein} E.,  {Springel}
  V.,   {Moscardini} L.,  2014, \mn@doi [\mnras] {10.1093/mnras/stu259}, \href
  {http://adsabs.harvard.edu/abs/2014MNRAS.440...75B} {440, 75}

\bibitem[\protect\citeauthoryear{{Barreira}, {Li}, {Hellwing}, {Lombriser},
  {Baugh}  \& {Pascoli}}{{Barreira} et~al.}{2014}]{2014JCAP...04..029B}
{Barreira} A.,  {Li} B.,  {Hellwing} W.~A.,  {Lombriser} L.,  {Baugh} C.~M.,
  {Pascoli} S.,  2014, \mn@doi [\jcap] {10.1088/1475-7516/2014/04/029}, \href
  {http://adsabs.harvard.edu/abs/2014JCAP...04..029B} {4, 029}

\bibitem[\protect\citeauthoryear{Battaglia}{Battaglia}{2016}]{Battaglia:2016xbi}
Battaglia N.,  2016, \mn@doi [JCAP] {10.1088/1475-7516/2016/08/058}, 1608, 058

\bibitem[\protect\citeauthoryear{Berti et~al.}{Berti
  et~al.}{2015}]{Berti:2015itd}
Berti E.,  et~al., 2015, \mn@doi [Class. Quant. Grav.]
  {10.1088/0264-9381/32/24/243001}, 32, 243001

\bibitem[\protect\citeauthoryear{Bertotti, Iess  \& Tortora}{Bertotti
  et~al.}{2003}]{Bertotti:2003rm}
Bertotti B.,  Iess L.,   Tortora P.,  2003, \mn@doi [Nature]
  {10.1038/nature01997}, 425, 374

\bibitem[\protect\citeauthoryear{{Bhattacharya} \& {Kosowsky}}{{Bhattacharya}
  \& {Kosowsky}}{2008}]{2008JCAP...08..030B}
{Bhattacharya} S.,  {Kosowsky} A.,  2008, \mn@doi [\jcap]
  {10.1088/1475-7516/2008/08/030}, \href
  {http://adsabs.harvard.edu/abs/2008JCAP...08..030B} {8, 030}

\bibitem[\protect\citeauthoryear{Bonvin}{Bonvin}{2008}]{Bonvin:2008ni}
Bonvin C.,  2008, \mn@doi [Phys. Rev.] {10.1103/PhysRevD.78.123530}, D78,
  123530

\bibitem[\protect\citeauthoryear{Bonvin \& Durrer}{Bonvin \&
  Durrer}{2011}]{Bonvin:2011bg}
Bonvin C.,  Durrer R.,  2011, \mn@doi [Phys. Rev.]
  {10.1103/PhysRevD.84.063505}, D84, 063505

\bibitem[\protect\citeauthoryear{Bonvin \& Fleury}{Bonvin \&
  Fleury}{2018}]{Bonvin:2018ckp}
Bonvin C.,  Fleury P.,  2018, \mn@doi [JCAP] {10.1088/1475-7516/2018/05/061},
  1805, 061

\bibitem[\protect\citeauthoryear{Bonvin, Hui  \& Gaztanaga}{Bonvin
  et~al.}{2014}]{Bonvin:2013ogt}
Bonvin C.,  Hui L.,   Gaztanaga E.,  2014, \mn@doi [Phys. Rev.]
  {10.1103/PhysRevD.89.083535}, D89, 083535

\bibitem[\protect\citeauthoryear{Bonvin, Andrianomena, Bacon, Clarkson,
  Maartens, Moloi  \& Bull}{Bonvin et~al.}{2017}]{Bonvin:2016dze}
Bonvin C.,  Andrianomena S.,  Bacon D.,  Clarkson C.,  Maartens R.,  Moloi T.,
   Bull P.,  2017, \mn@doi [Mon. Not. Roy. Astron. Soc.]
  {10.1093/mnras/stx2049}, 472, 3936

\bibitem[\protect\citeauthoryear{Bull}{Bull}{2016}]{Bull:2015lja}
Bull P.,  2016, \mn@doi [Astrophys. J.] {10.3847/0004-637X/817/1/26}, 817, 26

\bibitem[\protect\citeauthoryear{Burrage \& Sakstein}{Burrage \&
  Sakstein}{2016}]{Burrage:2016bwy}
Burrage C.,  Sakstein J.,  2016, \mn@doi [JCAP]
  {10.1088/1475-7516/2016/11/045}, 1611, 045

\bibitem[\protect\citeauthoryear{Casaponsa, Heavens, Kitching, Miller, Barreiro
   \& Martinez-Gonzalez}{Casaponsa et~al.}{2013}]{Casaponsa:2012tq}
Casaponsa B.,  Heavens A.~F.,  Kitching T.~D.,  Miller L.,  Barreiro R.~B.,
  Martinez-Gonzalez E.,  2013, \mn@doi [Mon. Not. Roy. Astron. Soc.]
  {10.1093/mnras/stt088}, 430, 2844

\bibitem[\protect\citeauthoryear{Challinor \& Lewis}{Challinor \&
  Lewis}{2011}]{Challinor:2011bk}
Challinor A.,  Lewis A.,  2011, \mn@doi [Phys.Rev.]
  {10.1103/PhysRevD.84.043516}, D84, 043516

\bibitem[\protect\citeauthoryear{{DESI Collaboration} et~al.,}{{DESI
  Collaboration} et~al.}{2016}]{2016arXiv161100036D}
{DESI Collaboration} et~al., 2016, preprint, \href
  {http://adsabs.harvard.edu/abs/2016arXiv161100036D} {} (\mn@eprint {arXiv}
  {1611.00036})

\bibitem[\protect\citeauthoryear{Damour \& Taylor}{Damour \&
  Taylor}{1992}]{Damour:1991rd}
Damour T.,  Taylor J.~H.,  1992, \mn@doi [Phys. Rev.]
  {10.1103/PhysRevD.45.1840}, D45, 1840

\bibitem[\protect\citeauthoryear{{Desjacques}, {Jeong}  \&
  {Schmidt}}{{Desjacques} et~al.}{2016}]{2016arXiv161109787D}
{Desjacques} V.,  {Jeong} D.,   {Schmidt} F.,  2016, preprint, \href
  {http://adsabs.harvard.edu/abs/2016arXiv161109787D} {} (\mn@eprint {arXiv}
  {1611.09787})

\bibitem[\protect\citeauthoryear{{Djorgovski} \& {Davis}}{{Djorgovski} \&
  {Davis}}{1987}]{1987ApJ...313...59D}
{Djorgovski} S.,  {Davis} M.,  1987, \mn@doi [\apj] {10.1086/164948}, \href
  {https://ui.adsabs.harvard.edu/abs/1987ApJ...313...59D} {313, 59}

\bibitem[\protect\citeauthoryear{Esposito-Farese}{Esposito-Farese}{1996}]{EspositoFarese:1996si}
Esposito-Farese G.,  1996, in {Colloquium on Pulsar Timing, General Relativity,
  and the Internal Structure of Neutron Stars Amsterdam, Netherlands, September
  24-28, 1996}. pp 90--6984 (\mn@eprint {arXiv} {gr-qc/9612039})

\bibitem[\protect\citeauthoryear{Everitt et~al.}{Everitt
  et~al.}{2011}]{Everitt:2011hp}
Everitt C. W.~F.,  et~al., 2011, \mn@doi [Phys. Rev. Lett.]
  {10.1103/PhysRevLett.106.221101}, 106, 221101

\bibitem[\protect\citeauthoryear{Faber \& Jackson}{Faber \&
  Jackson}{1976}]{Faber:1976sn}
Faber S.~M.,  Jackson R.~E.,  1976, \mn@doi [Astrophys. J.] {10.1086/154215},
  204, 668

\bibitem[\protect\citeauthoryear{Giannantonio, Martinelli, Silvestri  \&
  Melchiorri}{Giannantonio et~al.}{2010}]{giannantonio2010new}
Giannantonio T.,  Martinelli M.,  Silvestri A.,   Melchiorri A.,  2010, Journal
  of Cosmology and Astroparticle Physics, 2010, 030

\bibitem[\protect\citeauthoryear{Gleyzes, Langlois, Piazza  \&
  Vernizzi}{Gleyzes et~al.}{2013}]{Gleyzes:2013ooa}
Gleyzes J.,  Langlois D.,  Piazza F.,   Vernizzi F.,  2013, \mn@doi [JCAP]
  {10.1088/1475-7516/2013/08/025}, 1308, 025

\bibitem[\protect\citeauthoryear{Gleyzes, Langlois, Mancarella  \&
  Vernizzi}{Gleyzes et~al.}{2015}]{Gleyzes:2015pma}
Gleyzes J.,  Langlois D.,  Mancarella M.,   Vernizzi F.,  2015, \mn@doi [JCAP]
  {10.1088/1475-7516/2015/08/054}, 1508, 054

\bibitem[\protect\citeauthoryear{{Gronke}, {Llinares}, {Mota}  \&
  {Winther}}{{Gronke} et~al.}{2015}]{2015MNRAS.449.2837G}
{Gronke} M.,  {Llinares} C.,  {Mota} D.~F.,   {Winther} H.~A.,  2015, \mn@doi
  [\mnras] {10.1093/mnras/stv496}, \href
  {http://adsabs.harvard.edu/abs/2015MNRAS.449.2837G} {449, 2837}

\bibitem[\protect\citeauthoryear{Hall, Bonvin  \& Challinor}{Hall
  et~al.}{2013}]{Hall:2012wd}
Hall A.,  Bonvin C.,   Challinor A.,  2013, \mn@doi [Phys. Rev.]
  {10.1103/PhysRevD.87.064026}, D87, 064026

\bibitem[\protect\citeauthoryear{Heavens, Alsing  \& Jaffe}{Heavens
  et~al.}{2013}]{Heavens:2013gol}
Heavens A.,  Alsing J.,   Jaffe A.,  2013, \mn@doi [Mon. Not. Roy. Astron.
  Soc.] {10.1093/mnrasl/slt045}, 433, 6

\bibitem[\protect\citeauthoryear{{Hellwing}, {Barreira}, {Frenk}, {Li}  \&
  {Cole}}{{Hellwing} et~al.}{2014}]{2014PhRvL.112v1102H}
{Hellwing} W.~A.,  {Barreira} A.,  {Frenk} C.~S.,  {Li} B.,   {Cole} S.,  2014,
  \mn@doi [Physical Review Letters] {10.1103/PhysRevLett.112.221102}, \href
  {http://adsabs.harvard.edu/abs/2014PhRvL.112v1102H} {112, 221102}

\bibitem[\protect\citeauthoryear{Hu, Liguori, Bartolo  \& Matarrese}{Hu
  et~al.}{2013}]{Hu:2013aqa}
Hu B.,  Liguori M.,  Bartolo N.,   Matarrese S.,  2013, \mn@doi [Phys. Rev.]
  {10.1103/PhysRevD.88.123514}, D88, 123514

\bibitem[\protect\citeauthoryear{{Hudson} \& {Turnbull}}{{Hudson} \&
  {Turnbull}}{2012}]{2012ApJ...751L..30H}
{Hudson} M.~J.,  {Turnbull} S.~J.,  2012, \mn@doi [\apjl]
  {10.1088/2041-8205/751/2/L30}, \href
  {http://adsabs.harvard.edu/abs/2012ApJ...751L..30H} {751, L30}

\bibitem[\protect\citeauthoryear{{Ivarsen}, {Bull}, {Llinares}  \&
  {Mota}}{{Ivarsen} et~al.}{2016}]{2016A&A...595A..40I}
{Ivarsen} M.~F.,  {Bull} P.,  {Llinares} C.,   {Mota} D.,  2016, \mn@doi [\aap]
  {10.1051/0004-6361/201628604}, \href
  {http://adsabs.harvard.edu/abs/2016A%26A...595A..40I} {595, A40}

\bibitem[\protect\citeauthoryear{{Johnson}, {Blake}, {Dossett}, {Koda},
  {Parkinson}  \& {Joudaki}}{{Johnson} et~al.}{2016}]{2016MNRAS.458.2725J}
{Johnson} A.,  {Blake} C.,  {Dossett} J.,  {Koda} J.,  {Parkinson} D.,
  {Joudaki} S.,  2016, \mn@doi [\mnras] {10.1093/mnras/stw447}, \href
  {http://adsabs.harvard.edu/abs/2016MNRAS.458.2725J} {458, 2725}

\bibitem[\protect\citeauthoryear{Kaiser \& Hudson}{Kaiser \&
  Hudson}{2015}]{Kaiser:2014jca}
Kaiser N.,  Hudson M.~J.,  2015, \mn@doi [Mon. Not. Roy. Astron. Soc.]
  {10.1093/mnras/stv693}, 450, 883

\bibitem[\protect\citeauthoryear{Kashlinsky, Atrio-Barandela, Kocevski  \&
  Ebeling}{Kashlinsky et~al.}{2009}]{Kashlinsky:2008ut}
Kashlinsky A.,  Atrio-Barandela F.,  Kocevski D.,   Ebeling H.,  2009, \mn@doi
  [Astrophys. J.] {10.1086/592947}, 686, L49

\bibitem[\protect\citeauthoryear{{Koyama}, {Maartens}  \& {Song}}{{Koyama}
  et~al.}{2009}]{2009JCAP...10..017K}
{Koyama} K.,  {Maartens} R.,   {Song} Y.-S.,  2009, \mn@doi [\jcap]
  {10.1088/1475-7516/2009/10/017}, \href
  {http://adsabs.harvard.edu/abs/2009JCAP...10..017K} {10, 017}

\bibitem[\protect\citeauthoryear{Kramer et~al.,}{Kramer
  et~al.}{2006}]{kramer2006tests}
Kramer M.,  et~al., 2006, Science, 314, 97

\bibitem[\protect\citeauthoryear{Lagos, Bellini, Noller, Ferreira  \&
  Baker}{Lagos et~al.}{2018}]{Lagos:2017hdr}
Lagos M.,  Bellini E.,  Noller J.,  Ferreira P.~G.,   Baker T.,  2018, \mn@doi
  [JCAP] {10.1088/1475-7516/2018/03/021}, 1803, 021

\bibitem[\protect\citeauthoryear{{Leonard}, {Ferreira}  \& {Heymans}}{{Leonard}
  et~al.}{2015}]{2015JCAP...12..051L}
{Leonard} C.~D.,  {Ferreira} P.~G.,   {Heymans} C.,  2015, \mn@doi [\jcap]
  {10.1088/1475-7516/2015/12/051}, \href
  {http://adsabs.harvard.edu/abs/2015JCAP...12..051L} {12, 051}

\bibitem[\protect\citeauthoryear{{Macaulay}, {Feldman}, {Ferreira}, {Hudson}
  \& {Watkins}}{{Macaulay} et~al.}{2011}]{2011MNRAS.414..621M}
{Macaulay} E.,  {Feldman} H.,  {Ferreira} P.~G.,  {Hudson} M.~J.,   {Watkins}
  R.,  2011, \mn@doi [\mnras] {10.1111/j.1365-2966.2011.18426.x}, \href
  {http://adsabs.harvard.edu/abs/2011MNRAS.414..621M} {414, 621}

\bibitem[\protect\citeauthoryear{Montanari \& Durrer}{Montanari \&
  Durrer}{2012}]{Montanari:2012me}
Montanari F.,  Durrer R.,  2012, \mn@doi [Phys. Rev.]
  {10.1103/PhysRevD.86.063503}, D86, 063503

\bibitem[\protect\citeauthoryear{Moradinezhad~Dizgah \&
  Durrer}{Moradinezhad~Dizgah \& Durrer}{2016}]{Dizgah:2016bgm}
Moradinezhad~Dizgah A.,  Durrer R.,  2016, \mn@doi [JCAP]
  {10.1088/1475-7516/2016/09/035}, 1609, 035

\bibitem[\protect\citeauthoryear{{Mueller}, {de Bernardis}, {Bean}  \&
  {Niemack}}{{Mueller} et~al.}{2015}]{2015ApJ...808...47M}
{Mueller} E.-M.,  {de Bernardis} F.,  {Bean} R.,   {Niemack} M.~D.,  2015,
  \mn@doi [\apj] {10.1088/0004-637X/808/1/47}, \href
  {http://adsabs.harvard.edu/abs/2015ApJ...808...47M} {808, 47}

\bibitem[\protect\citeauthoryear{Papai \& Szapudi}{Papai \&
  Szapudi}{2008}]{Papai:2008bd}
Papai P.,  Szapudi I.,  2008, \mn@doi [Mon. Not. Roy. Astron. Soc.]
  {10.1111/j.1365-2966.2008.13572.x}, 389, 292

\bibitem[\protect\citeauthoryear{{Planck Collaboration}}{{Planck
  Collaboration}}{2018}]{2018arXiv180706209P}
{Planck Collaboration} 2018, preprint, \href
  {http://adsabs.harvard.edu/abs/2018arXiv180706209P} {} (\mn@eprint {arXiv}
  {1807.06209})

\bibitem[\protect\citeauthoryear{{{Planck Collaboration XIII}}}{{{Planck
  Collaboration XIII}}}{2016}]{Ade:2015xua}
{{Planck Collaboration XIII}} 2016, \mn@doi [Astron. Astrophys.]
  {10.1051/0004-6361/201525830}, 594, A13

\bibitem[\protect\citeauthoryear{{{Planck Collaboration XIV}}}{{{Planck
  Collaboration XIV}}}{2016}]{Ade:2015rim}
{{Planck Collaboration XIV}} 2016, \mn@doi [Astron. Astrophys.]
  {10.1051/0004-6361/201525814}, 594, A14

\bibitem[\protect\citeauthoryear{Pogosian, Silvestri, Koyama  \& Zhao}{Pogosian
  et~al.}{2010}]{Pogosian:2010tj}
Pogosian L.,  Silvestri A.,  Koyama K.,   Zhao G.-B.,  2010, \mn@doi [Phys.
  Rev.] {10.1103/PhysRevD.81.104023}, D81, 104023

\bibitem[\protect\citeauthoryear{{Reyes}, {Mandelbaum}, {Seljak}, {Baldauf},
  {Gunn}, {Lombriser}  \& {Smith}}{{Reyes} et~al.}{2010}]{2010Natur.464..256R}
{Reyes} R.,  {Mandelbaum} R.,  {Seljak} U.,  {Baldauf} T.,  {Gunn} J.~E.,
  {Lombriser} L.,   {Smith} R.~E.,  2010, \mn@doi [\nat] {10.1038/nature08857},
  \href {http://adsabs.harvard.edu/abs/2010Natur.464..256R} {464, 256}

\bibitem[\protect\citeauthoryear{Sakstein}{Sakstein}{2018}]{Sakstein:2017pqi}
Sakstein J.,  2018, \mn@doi [Phys. Rev.] {10.1103/PhysRevD.97.064028}, D97,
  064028

\bibitem[\protect\citeauthoryear{Samushia et~al.}{Samushia
  et~al.}{2014}]{Samushia:2013yga}
Samushia L.,  et~al., 2014, \mn@doi [Mon. Not. Roy. Astron. Soc.]
  {10.1093/mnras/stu197}, 439, 3504

\bibitem[\protect\citeauthoryear{Schmidt, Leauthaud, Massey, Rhodes, George,
  Koekemoer, Finoguenov  \& Tanaka}{Schmidt et~al.}{2012}]{Schmidt:2011qj}
Schmidt F.,  Leauthaud A.,  Massey R.,  Rhodes J.,  George M.~R.,  Koekemoer
  A.~M.,  Finoguenov A.,   Tanaka M.,  2012, \mn@doi [Astrophys. J.]
  {10.1088/2041-8205/744/2/L22}, 744, L22

\bibitem[\protect\citeauthoryear{Song, Hu  \& Sawicki}{Song
  et~al.}{2007a}]{song2007large}
Song Y.-S.,  Hu W.,   Sawicki I.,  2007a, Physical Review D, 75, 044004

\bibitem[\protect\citeauthoryear{Song, Peiris  \& Hu}{Song
  et~al.}{2007b}]{song2007cosmological}
Song Y.-S.,  Peiris H.,   Hu W.,  2007b, Physical Review D, 76, 063517

\bibitem[\protect\citeauthoryear{Szalay, Matsubara  \& Landy}{Szalay
  et~al.}{1998}]{Szalay:1997cc}
Szalay A.~S.,  Matsubara T.,   Landy S.~D.,  1998, \mn@doi [Astrophys. J.]
  {10.1086/311293}, 498, L1

\bibitem[\protect\citeauthoryear{Szapudi}{Szapudi}{2004}]{Szapudi:2004gh}
Szapudi I.,  2004, \mn@doi [Astrophys. J.] {10.1086/423168}, 614, 51

\bibitem[\protect\citeauthoryear{Tansella, Jelic-Cizmek, Bonvin  \&
  Durrer}{Tansella et~al.}{2018}]{tansella2018coffe}
Tansella V.,  Jelic-Cizmek G.,  Bonvin C.,   Durrer R.,  2018, arXiv preprint
  arXiv:1806.11090

\bibitem[\protect\citeauthoryear{Taylor \& Weisberg}{Taylor \&
  Weisberg}{1982}]{Taylor:1982zz}
Taylor J.~H.,  Weisberg J.~M.,  1982, \mn@doi [Astrophys. J.] {10.1086/159690},
  253, 908

\bibitem[\protect\citeauthoryear{{Taylor}, {Wolszczan}, {Damour}  \&
  {Weisberg}}{{Taylor} et~al.}{1992}]{Taylor:1991yt}
{Taylor} J.~H.,  {Wolszczan} A.,  {Damour} T.,   {Weisberg} J.~M.,  1992,
  \mn@doi [\nat] {10.1038/355132a0}, \href
  {http://adsabs.harvard.edu/abs/1992Natur.355..132T} {355, 132}

\bibitem[\protect\citeauthoryear{Touboul et~al.}{Touboul
  et~al.}{2017}]{Touboul:2017grn}
Touboul P.,  et~al., 2017, \mn@doi [Phys. Rev. Lett.]
  {10.1103/PhysRevLett.119.231101}, 119, 231101

\bibitem[\protect\citeauthoryear{{Tully} \& {Fisher}}{{Tully} \&
  {Fisher}}{1977}]{1977A&A....54..661T}
{Tully} R.~B.,  {Fisher} J.~R.,  1977, \aap, \href
  {http://adsabs.harvard.edu/abs/1977A%26A....54..661T} {54, 661}

\bibitem[\protect\citeauthoryear{{Wang}, {Koribalski}, {Serra}, {van der
  Hulst}, {Roychowdhury}, {Kamphuis}  \& {Chengalur}}{{Wang}
  et~al.}{2016}]{2016MNRAS.460.2143W}
{Wang} J.,  {Koribalski} B.~S.,  {Serra} P.,  {van der Hulst} T.,
  {Roychowdhury} S.,  {Kamphuis} P.,   {Chengalur} J.~N.,  2016, \mn@doi
  [\mnras] {10.1093/mnras/stw1099}, \href
  {http://adsabs.harvard.edu/abs/2016MNRAS.460.2143W} {460, 2143}

\bibitem[\protect\citeauthoryear{{Watkins}, {Feldman}  \& {Hudson}}{{Watkins}
  et~al.}{2009}]{2009MNRAS.392..743W}
{Watkins} R.,  {Feldman} H.~A.,   {Hudson} M.~J.,  2009, \mn@doi [\mnras]
  {10.1111/j.1365-2966.2008.14089.x}, \href
  {http://adsabs.harvard.edu/abs/2009MNRAS.392..743W} {392, 743}

\bibitem[\protect\citeauthoryear{{Wechsler} \& {DESI Collaboration}}{{Wechsler}
  \& {DESI Collaboration}}{2015}]{2015DESI}
{Wechsler} R.~H.,  {DESI Collaboration} 2015, in American Astronomical Society
  Meeting Abstracts \#225. p. 336.06

\bibitem[\protect\citeauthoryear{Weisberg \& Taylor}{Weisberg \&
  Taylor}{2005}]{Weisberg:2004hi}
Weisberg J.~M.,  Taylor J.~H.,  2005, ASP Conf. Ser., 328, 25

\bibitem[\protect\citeauthoryear{{Wex}}{{Wex}}{2014}]{Wex:2014nva}
{Wex} N.,  2014, preprint, \href
  {http://adsabs.harvard.edu/abs/2014arXiv1402.5594W} {} (\mn@eprint {arXiv}
  {1402.5594})

\bibitem[\protect\citeauthoryear{Will}{Will}{2006}]{Will:2005va}
Will C.~M.,  2006, \mn@doi [Living Rev. Rel.] {10.12942/lrr-2006-3}, 9, 3

\bibitem[\protect\citeauthoryear{{Will}}{{Will}}{2011}]{2011PNAS..108.5938W}
{Will} C.~M.,  2011, \mn@doi [Proceedings of the National Academy of Science]
  {10.1073/pnas.1103127108}, \href
  {http://adsabs.harvard.edu/abs/2011PNAS..108.5938W} {108, 5938}

\bibitem[\protect\citeauthoryear{Williams, Turyshev  \& Boggs}{Williams
  et~al.}{2004}]{Williams:2004qba}
Williams J.~G.,  Turyshev S.~G.,   Boggs D.~H.,  2004, \mn@doi [Phys. Rev.
  Lett.] {10.1103/PhysRevLett.93.261101}, 93, 261101

\bibitem[\protect\citeauthoryear{Yoo, Fitzpatrick  \& Zaldarriaga}{Yoo
  et~al.}{2009}]{Yoo:2009au}
Yoo J.,  Fitzpatrick A.~L.,   Zaldarriaga M.,  2009, \mn@doi [Phys.Rev.]
  {10.1103/PhysRevD.80.083514}, D80, 083514

\bibitem[\protect\citeauthoryear{{Zhang}, {Liguori}, {Bean}  \&
  {Dodelson}}{{Zhang} et~al.}{2007}]{2007PhRvL..99n1302Z}
{Zhang} P.,  {Liguori} M.,  {Bean} R.,   {Dodelson} S.,  2007, \mn@doi
  [Physical Review Letters] {10.1103/PhysRevLett.99.141302}, \href
  {http://adsabs.harvard.edu/abs/2007PhRvL..99n1302Z} {99, 141302}

\bibitem[\protect\citeauthoryear{{Zheng}, {Zhang}  \& {Jing}}{{Zheng}
  et~al.}{2015}]{2015PhRvD..91l3512Z}
{Zheng} Y.,  {Zhang} P.,   {Jing} Y.,  2015, \mn@doi [\prd]
  {10.1103/PhysRevD.91.123512}, \href
  {http://adsabs.harvard.edu/abs/2015PhRvD..91l3512Z} {91, 123512}

\makeatother
\end{thebibliography}



\bsp	
\label{lastpage}
\end{document}